\documentclass[12pt]{article}
\pdfoutput=1

\usepackage{amsmath,amsthm,amssymb,euscript,epsf,epsfig,cite,tikz}
\usepackage{array,longtable,fancybox,comment,tikz-cd,rotating,pbox,hyperref} 
\usepackage[vcentermath]{youngtab}
\usepackage{empheq}
\usetikzlibrary{shapes,arrows,calc,decorations.markings}

\newcommand{\boxedeq}[2]{\begin{empheq}[box={\fboxsep=6pt\fbox}]{align}\label{#1}#2\end{empheq}}

\makeatletter
\@addtoreset{equation}{section}
\makeatother

\def\one{{\hbox{ 1\kern-.8mm l}}}
\def\zero{{\hbox{ 0\kern-1.5mm 0}}}

\def\mC{ \mathbb{C}}

 \def\cN{{\cal N}} \def\cO{{\cal O}}
  \def\cR{{\cal R}}
\def\cS{{\cal S}}  
\def\cV{{\cal V}}  
 
\def\mg{ \mathfrak{g}}

\def\psu{ {\rm{psu} } }

\newcommand{\be}{\begin{equation}}
\newcommand{\ee}{\end{equation}}
\newcommand{\beq}{\begin{equation}}
\newcommand{\eeq}{\end{equation}}
\newcommand{\bea}{\begin{eqnarray}\displaystyle}
\newcommand{\eea}{\end{eqnarray}}

\newcommand{\Tr}{{\rm Tr}}
\newcommand{\tr}{{\rm tr}}
 
\newcommand{\mR}{\mathbb{R}} 

\begin{document}

\begin{flushright}
QMUL-PH-22-34
\end{flushright}

\medskip

\begin{center}

{\Large \bf  $\mathcal{N}=4$ SYM, (super)-polynomial rings and emergent quantum mechanical symmetries.}

\bigskip
Robert de Mello Koch$^{a,c,{\tiny \yng(2)}}$ and  Sanjaye Ramgoolam$^{b,c,{\tiny\yng(1,1)}}$ 

\bigskip

{\small
$^{a}${\em School of Science, Huzhou University, Huzhou 313000, China,}\\
\medskip
$^{b}$
{\em Centre for Theoretical Physics, School of Physical and Chemical Sciences } \\
{\em Queen Mary University of London, London E1 4NS, United Kingdom }\\
\medskip
$^{c}$
{\em  School of Physics and Mandelstam Institute for Theoretical Physics,}
{\em University of Witwatersrand, Wits, 2050, South Africa} \\
}

\begin{abstract}

The structure of half-BPS representations of  psu$(2,2|4)$ leads to the definition of a super-polynomial ring $\mathcal{R}(8|8)$ which admits a realisation of  psu$(2,2|4)$ in terms of differential operators on the super-ring.
The character of the half-BPS fundamental field representation encodes the resolution of the  representation in terms of an  exact sequence of modules of $\mathcal{R}(8|8)$. The half-BPS representation is realized by quotienting the super-ring by a quadratic ideal, equivalently by setting to zero certain quadratic polynomials in the generators of the super-ring. This description of the half-BPS fundamental field irreducible representation of psu$(2,2|4)$ in terms of a super-polynomial ring is an example of a more general construction of  lowest-weight representations of Lie (super-) algebras using polynomial rings generated by a commuting subspace of the standard raising operators, corresponding to positive roots of the Lie (super-) algebra. We illustrate the construction using simple examples of representations of su(3) and su(4). These results lead to the definition of  a notion of quantum mechanical emergence for oscillator realisations of symmetries,  which is based on ideals in the ring of polynomials in the creation operators.

\end{abstract}
\end{center}

\vfill

\noindent 
{\small{ 
\noindent 
 {E-mails:  $^{\tiny\yng(2)}$robert@zjhu.edu.cn,${}^{\tiny\yng(1,1)}$ s.ramgoolam@qmul.ac.uk} 
}}

\newpage 

\tableofcontents

\section{Introduction}

The AdS/CFT correspondence \cite{malda,GKP,witten} between $\cN =4 $ super Yang-Mills (SYM) theory with $U(N)$ gauge group and string theory on $\rm{AdS}_5\times \rm{S}^5$ has led to  new algebraic perspectives on the quantum states and correlators in SYM theory aimed at uncovering the mechanisms of emergence of stringy branes and geometries from SYM. Giant gravitons \cite{mst} were recognised at an early stage as important half-BPS quantum states whose properties are highly sensitive to non-perturbative finite $N$ effects. General half-BPS states in the CFT side correspond to holomorphic gauge invariant functions of a complex matrix  $Z$ \cite{Eden2000NPB,Eden2000PLB}.
Sub-determinants of $Z$ were identified with a class of giant graviton states \cite{BBNS}.  This leads to a  classification of general  half-BPS states in terms of Young diagrams \cite{CJR}  and  allows a natural identification of CFT duals of single and multi-giant states.  The Young diagram classification and the associated correspondence with  free fermions \cite{CJR,Ber2004} were also shown to appear in the classification of half-BPS super-gravity solutions \cite{LLM}.

Excitations of giant graviton states in AdS are described from the CFT point of view by using perturbations of the Young diagram states, which arise by insertions of impurity matrices \cite{BMLN2002,BBFH2005,DSS2007}. This has led to a detailed treatment of multi-matrix invariants and associated Hilbert spaces, motivated by  the AdS/CFT correspondence.  Representation theoretic bases of multi-matrix invariants, orthogonal in the free field inner product arising from SYM, were found \cite{KR2007,BHR2007,DBC2008,DBS2008,BHR2008}. At weak coupling elements of these bases mix weakly allowing an evaluation of the spectrum of one loop anomalous dimensions \cite{Carlson:2011hy,deMelloKoch:2011wah,deMelloKoch:2011ci,deMelloKoch:2012ck,Berenstein:2013md}. An  underlying algebraic structure for these bases,  of associative algebras and two-dimensional topological quantum field theories (TQFT2), related to permutations and more generally to diagram algebras, was described \cite{EHS,PCA,QuivCalc,KimuraTFT1,KimuraTFT2}. Essentially, the representation theoretic bases arise from Fourier transforms on the algebras. These algebras are equipped with non-degenerate pairings arising from the free SYM theory, which leads to the links with TQFT2.

The free field inner products of composite operators, e.g. operators in the half-BPS sector or more general multi-matrix operators,  generically take the form 
\bea 
\langle \cO_{ A } ( x_1 ) \cO_{ B } ( x_2 ) \rangle 
=  { M_{ A B }  \over ( x_1 - x_2 )^{ 2 \Delta } } 
  \eea
where $\cO_{A} , \cO_{ B } $ run over a complete basis for operators of the same dimension $\Delta $. The combinatorial diagonalisation problems involve finding a basis for the operators of any fixed dimension which  diagonalises the matrix $M_{ AB }$. Given that the structure of $M_{AB}$ and its diagonalisation  are captured by TQFT2, in  multi-matrix problems of interest in AdS/CFT,  it is natural to turn to ask if TQFT2 also captures the space-time dependence of two-point functions. This motivated the introduction of so$(4,2)$ equivariant TQFT2 in \cite{CFT4TFT2}, as part of the CFT4-TFT2 programme. This was used to concretely describe the state space and associative algebra underlying general correlators in free scalar field theory in four dimensions.  The algebraic approach of CFT4-TFT2 has been extended to describe free fermion CFTs and to give algorithms for general constructions of free field primaries in terms of polynomial rings \cite{deMelloKoch:2017dgi,deMelloKoch:2017caf,DeMelloKoch:2018hyq,PrimsCC,PCFTDiag}. 

The starting point in the CFT4-TFT2 approach for the free scalar is to consider the space of states in radial quantization as a space of polynomials in the momenta 

\bea 
 \lim_{ x \rightarrow 0 } \phi ( x ) |0 \rangle & \rightarrow &  |\phi \rangle = 1  |\phi \rangle  \cr 
  \lim_{ x \rightarrow 0 } ( \partial_{ \mu} \phi ( x ) ) 
 |0 \rangle & \rightarrow &  P_{ \mu } |\phi \rangle \cr 
 & \vdots &
\eea
The space of polynomials carries a representation of so$(4,2) $ which is given explicitly in terms of differential operators. In this realization the momentum operators  $P_{\mu}$ act by multiplication on the polynomials, while the scaling dimension  operator  $D$, special conformal transformations $K_{\mu}$ and Lorentz rotations $M_{\mu\nu}$ are given by 
\bea 
D&=&P_{\mu}{\partial\over\partial P_{\mu}} +1 \cr 
K_{\mu}&=&P_{\mu}{\partial\over\partial P_{\alpha}}{\partial\over\partial P_{\alpha}}-  
2{\partial\over\partial P_{\mu}}P_{\alpha}{\partial\over\partial P_{\alpha}}\cr 
M_{\mu\nu}&=&P_{\mu}{\partial\over\partial P_{\nu}}-P_{\nu} {\partial\over\partial_{P_\mu}}
\eea
The so$(4,2)$ commutation relations are easily verified, and this representation was useful as a starting point for describing a deformed co-product related to anomalous dimensions in  \cite{PCFTDiag}.  This can also be viewed as an oscillator realization by writing $P_\mu \rightarrow a_{\mu}^{\dagger}$ and  ${\partial\over\partial P^{\mu}}\rightarrow a_\mu$. 

In this paper, we initiate the development of  the CFT4-TFT2 programme for U$(N)$ $\cN=4$ SYM theory. The first task is to describe a super-oscillator realisation of the super-conformal algebra generalising the free scalar so$(4,2)$ realisation, which describes the half-BPS multiplet built from $\tr Z^n $, for any positive integer $n$, in terms of a super-polynomial ring. A case of special interest is $n=1$, i.e. the multiplet containing $\tr Z $. In the $U(1)$ case, this gives  the irreducible representation (irrep) of psu$(2,2|4)$ describing the fundamental field representation of U$(1)$ super Yang-Mills theory. The structure of these half-BPS representations turns out to reveal an interesting story. 

We show that the approach of rings and quotients works equally well in describing the physical state space in full $\cN=4$ super Yang-Mills theory. The half-BPS representations have a lowest weight state which is annihilated by all of the $S$ operators as well as eight of the $Q$ operators. This leaves eight of the $Q$'s which do not annihilate the BPS state. An elementary deduction from available formulae in the literature also shows  that eight bosonic generators of psu$(2,2|4)$ do not annihilate the ground state of the half-BPS representation. The fermionic generators anti-commute, the bosonic generators commute, while the commutators of the bosonic and fermionic generators vanish. This means that the 8 bosons and 8 fermions generate a graded-commutative or super-polynomial ring $\cR(8|8)$. 

A first important result in this paper is the description of the realization of psu$(2,2|4) $ in terms of  differential operators on the ring $\cR(8|8)$. While there are a number of super-oscillator realisations of psu$(2,2|4) $ (see \cite{Beisert:2003jj,Dobrev:2018lsx,wittenoscilattor}), as far as we are aware, the oscillator realisation we present is new. The differential operators depend on an arbitrary parameter $\Delta$. By setting $\Delta =n$ we obtain a realization relevant for both the multiplet built on the superconformal primary ${\rm Tr}(Z^n)$ in the U$(N)$ theory or on $Z^n$ in the U(1) theory. The half-BPS irreducible representation is obtained by imposing some constraints on the elements of the ring. The neat way to describe this mathematically is to describe give a resolution of the half-BPS representation in terms of an exact sequence of modules of the ring $\cR(8 |8)$. This has the form 
\bea\label{BGGsup}   
0\rightarrow {\cal V}^{(F)}\rightarrow\cdots\rightarrow {\cal V}^{(1)}\rightarrow {\cal V}^{(0)}= \cR( 8|8) \rightarrow V^{\rm{BPS}}_{\rm{phys}}\rightarrow 0 
\eea
Here $ V^{\rm{BPS}}_{\rm{phys}} $ is the physical state space of the half-BPS representation, ${\cal V}^{(0)}$ is obtained by acting with the elements of the ring $\cR(8 |8)$ on the half-BPS ground state, so as a graded vector space is isomorphic to $\cR( 8|8)$. For the case $n=1$, corresponding to the fundamental field rep of U(1), or for the multiplet with superconformal primary ${\rm Tr}(Z)$ in the U$(N)$ theory, we give a very explicit description of the ideal 
 ${\cal V}^{(1)}$ which describes the constraints.  It turns out that the ideal is quadratic in the generators of the ring. We show, using algorithms for super-rings from SageMath \cite{SageMathOnline}, that this description of the half-BPS fundamental field representation is compatible with the known character formula from \cite{BDHO}. This is our second main result. 

In the course of developing these results, we found it useful to consider simpler examples of the general story at play. The general story is based on the fact  that lowest weight representations of a Lie (super-)algebra $\mg$  are realised by decomposing $\mg=\mg^- \oplus\mg^0\oplus\mg^+$, where $\mg^0$ is the Cartan sub-algebra and $\mg^{\pm}$ correspond to positive and negative roots.  The lowest weight representation can then be described using a finite resolution, the BGG resolution \cite{BGG}. The resolution gives an exact sequence of U$(\mg^+)$ modules which has the lowest weight representation as the last non-trivial space in the sequence. The sequence is $U(\mg)$-equivariant.  We modify this standard Lie algebra discussion to a polynomial ring resolution by considering a commuting sub-algebra $\mg^+_c $ of the raising sub-algebra. This leads to a resolution which employs modules of the  polynomial (super-)ring U$(\mg^+_c)$. We illustrate the general framework through examples.

The super-polynomial ring $\cR(8|8)$  that we use to give a resolution of the fundamental field representation of psu$(2,2|4)$ naturally defines a Fock space associated to $8$ bosonic and $8$ fermionic oscillators. We  define a notion of emergence in quantum mechanical oscillator systems, based on ideals in the ring generated by the creation operators. The oscillator systems have a manifest homogeneous symmetry of type u$(n|m)$ which preserves the space of states with a fixed number of creation operators. The ring generated by the creation operators also has an inhomogeneous symmetry which does not preserve the number of oscillators, and the modules of the ring provide a resolution for an irreducible representation of this inhomogeneous symmetry. We refer to this inhomogeneous symmetry as an emergent symmetry from the oscillator system.    

The paper is organised as follows. In Section \ref{polyringsliealgebras} we review relevant  background and key results from the theory of rings and modules. In this Section the reader will find the explanation of how rings naturally arise from a given irreducible representation of a Lie algebra, how resolutions give a deeper understanding of the character of the irreducible representation and how these ideas extend to super-rings. The ring theory computations that need to be carried out, are implemented in the free open-source mathematics software system SageMath \cite{SageMathOnline}. We give samples of Sage code that can easily be modified to study other examples. In Section \ref{symmetryandring} the super-polynomial ring ${\cal V}_{\cR}^{(0)}=\mathcal{R}(8|8)$, with eight bosonic and eight fermionic generators, relevant for  half-BPS  representations of $\cN =4$ SYM,  is introduced. A realisation of psu$(2,2|4)$ in terms of differential operators acting on this ring, which obey the correct commutators, is given. To complete the description and motivate what is to follow, we then explain how elements of the ring correspond to the elementary fields of $\cN =4$ super Yang-Mills theory  in the case of $U(1)$ gauge group and we motivate the need for a resolution in terms of modules in describing the half-BPS representation $V_{\rm{phys}}^{\rm BPS}$ cleanly. Section \ref{polyringsandchars}
constructs the required resolution. A complete identification of the constraints that must be imposed is given and their accuracy is confirmed by recovering the known character formula for the half-BPS fundamental field representation \cite{BDHO}.  Section \ref{emergentsymm} starts with the observation that the super-polynomial ring used to give a resolution of the BPS representation naturally defines a Fock space associated to $8$ bosonic and $8$ fermionic oscillators. This Fock space enjoys a hidden u$(8|8)$ symmetry underlying the half-BPS representations of ${\cal N}=4$ super Yang-Mills theory. State are written as polynomials in the creation operators acting on the Fock vacuum. The hidden u$(8|8)$ symmetry is a homogeneous symmetry, only mixes states constructed from polynomials of a fixed degree. The psu$(2,2|4)$ symmetry of ${\cal N}=4$ Yang-Mills theory is realized as an emergent symmetry in this Fock space. This section motivates both symmetries: the hidden symmetry, such as the homogeneous u$(8|8)$ symmetry, is useful in giving a physical formulation of the problem of constructing the syzygies of the resolution, 
while the non-homogeneous emergent symmetry exists in the ring and in the terminal module of the polynomial (super-)ring resolution.   Section \ref{summaryanddirections} summarises our results and suggests avenues for further research. The Appendices spell out various conventions, some applications to fermionic Fock spaces and collects some sample Sage codes. There is a mathematica notebook available with the source code of this arXiv submission, which verifies that the generators of Section \ref{symmetryandring} do indeed close the correct algebra, given in Appendix \ref{psu224}.

\section{Polynomial rings and Lie algebra representations }\label{polyringsliealgebras}

The BGG resolution is a powerful tool used to understand the structure of finite dimensional irreducible representations of a Lie algebra $\mg$ \cite{BGG}. The resolution gives an exact sequence relating Verma modules of $U(\mg^+)$ and the irrep $V$ of $\mg$ in which we are interested. Here $U(\mg^+)$ denotes the universal enveloping algebra of the raising operators of $\mg$. The resulting resolution is $\mg$ equivariant. Once constructed the BGG resolution gives, for example, a very direct route to the celebrated Weyl character formula. The representation we are interested in is the half BPS representation of ${\cal N}=4$ super Yang-Mills theory. Since it is a BPS representation, certain elements of $\mg^+$ annihilate the lowest weight state of the representation. As we will discuss in Section \ref{symmetryandring}, it is a non-trivial fact that the set of raising operators that do not annihilate the lowest weight state, define a commuting subalgebra $\mg^+_c$. Consequently, a natural generalization of the BGG resolution, motivated by ${\cal N}=4$ super Yang-Mills theory, is a resolution in terms of modules of $U(\mg^+_c)$,  the universal enveloping algebra of $\mg^+_c$. $U(\mg^+_c)$ is a (super-)polynomial ring. This leads to a central idea in this paper: the use of (super-)polynomial rings as a way to study Lie algebra representations. 

To develop this idea, we need certain results from the theory of rings and modules. This includes exact sequences of modules of a ring, used to produce the resolution of a module. The resolution of a module which is an irreducible representation, is directly related to the character of the irreducible representation. In Section \ref{RingBackground} we review the required background. The ring theory computations that need to be carried out, are implemented in the free open-source mathematics software system SageMath. We give samples of Sage code that can easily be modified to study other examples. In Section \ref{RingsfromLieAlgebras} we explain how rings naturally arise from a given irreducible representation of a Lie algebra, and how the resolution of a module gives insight into the structure of the character of the irreducible representation. The generators of the ring are associated with a commuting subalgebra of the Lie algebra so that we are studying commutative rings. The resulting resolutions are not $\mg$-equivariant. The well known BGG resolution is an alternative resolution related to the same irreducible representation, which employs lowest weight Verma modules. We compare the resolution we have constructed to the BGG resolution in Section \ref{BGG}. We will see that the BGG resolution has a natural interpretation in terms of Weyl's character formula. Ultimately we are interested in studying the super Lie algebra psu$(2,2|4)$, relevant for ${\cal N}=4$ super Yang-Mills theory. In this case we need to extend our ideas to super-rings, which include Grassmann elements. We formulate this discussion in Section \ref{supering}.

\subsection{Rings, Modules and Resolutions}\label{RingBackground}

We will be making use of commutative polynomial rings. In this section we will give a short summary of some of the key points, extracted from the extended discussion in \cite{Cox}. Physical applications of this mathematics in the context of counting BPS operators can be found for example in \cite{Benvenuti:2006qr}. The notation $k[x_1,\cdots,x_n]$ denotes the collection of all polynomials in $x_1,\cdots , x_n$ with coefficients in the field $k$. Since polynomials in $k[x_1,\cdots,x_n]$ can be added and multiplied as usual, $k[x_1,\cdots,x_n]$ has the structure of a commutative ring. Further, since only non zero constant polynomials have multiplicative inverses in $k[x_1,\cdots,x_n]$, it is clear that $k[x_1,\cdots,x_n]$ is not a field. The degree of the polynomials provides a natural grading on the ring that we make use of below. We will make use of the notation
\bea
\langle f_1,\cdots,f_s\rangle\equiv\{p_1f_1+\cdots +p_sf_s : p_i \in k[x_1,\cdots,x_n]
\quad{\rm for}\quad i = 1, . . . , s\}
\eea
where $f_1,\cdots ,f_s \in k[x_1,\cdots , x_n]$ are a collection of polynomials.

An ideal ${\cal I}$ in the ring $k[x_1,\cdots ,x_n]$ is a non-empty subset such that $f + g \in{\cal I}$ whenever $f,g\in{\cal I}$ and $p f\in {\cal I}$ whenever $f \in{\cal I}$, and $p\in k[x_1,\cdots ,x_n]$ is an arbitrary polynomial. Given an ideal ${\cal I}$, we can construct the quotient ring $k[x_1,\cdots,x_n]/{\cal I}$ as follows: given $f \in k[x_1,\cdots, x_n]$, we have the coset
\bea
[f] = \{f + h : h \in {\cal I}\}
\eea
so that
\bea 
[f] = [g] \quad\Rightarrow\quad f-g \in {\cal I}
\eea
The quotient ring $k[x_1,\cdots ,x_n]/{\cal I}$ consists of all cosets $[f]$ for $f \in k[x_1,\cdots, x_n]$. These cosets inherit a sum and a product from the ring.

We will define modules over our polynomial ring $k[x_1,\cdots,x_n]$. Our interest in modules is because they are closely related to the representation theory of groups. Modules generalize vector spaces, by replacing the field of scalars by a ring: just as for a vector space, a module is an additive abelian group, and scalar multiplication is (i) distributive over the operation of addition between elements of the ring or module and (ii) is compatible with the ring multiplication. Clearly, modules are closely related to vector spaces. An important difference between them is that every vector space has a basis but not every module has a basis. Let $M$ be a module over a ring $R$. A set $F\subset M$ is a module basis for $M$ if and only if every $f\in M$ can be written in one and only one way as a linear combination
\bea
f = a_1 f_1 + \cdots + a_n f_n
\eea
where $a_i\in R$, and $f_i\in F$. A free module is a module that has a module basis i.e. a generating set consisting of linearly independent elements. Every vector space is a free module. If the ring of the coefficients is not a division ring then it's possible to construct non-free modules.

The concept of a module simultaneously generalizes the notions of an ideal and a quotient ring. An ideal ${\cal I}\subset R$ is an $R$-module, using the sum and product operations from $R$. The quotient ring $M = R/{\cal I}$ is an $R$-module under the quotient ring sum operation, and the scalar multiplication defined for cosets $[g] \in R/{\cal I}$ and $f\in R$ by $f[g] = [fg]\in R/{\cal I}$. The direct sum $M\oplus N$ is the set of ordered pairs $(f,g)$ with $f\in M$ and $g\in N$. $M\oplus N$ is an $R$-module under the component-wise sum and scalar multiplication operations. Let $F\subset M$ and let $N\subset M$ be the collection of all $f\in M$ of the form
\bea
f = a_1f_1 +\cdots + a_nf_n
\eea
with $a_i\in R$ and $f_i\in F$. $N$ is a submodule of $M$ called the submodule of $M$ generated by $F$. These submodules are natural generalizations of the ideals generated by given subsets of the ring $R$, so we use the same notation $\langle F\rangle$ for the submodule generated by set $F$. If $\langle F\rangle=M$, we say that $F$ spans (or generates) $M$. We say that $M$ is finitely generated if there is a finite set that generates $M$.

We can define the quotient of a module $M$ by a module $N$, denoted by $M/N$ as follows: If $N$ is a submodule of $M$, then the set of equivalence classes of elements of $M$ where $f,g \in M$ are equivalent if and only if $f-g\in N$ forms an $R$-module, $M/N$, with the operations induced from $M$.

We will also make use of module homomorphisms, which are the analogues of linear mappings between vector spaces. An $R$-module homomorphism between two $R$-modules
$M$ and $N$ is an $R$-linear map between $M$ and $N$, i.e. a map $\varphi :M \to N$ is an $R$-module homomorphism if for all $a\in R$ and all $f,g\in M$, we have
\bea
\varphi (af + g) = a\,\varphi (f) + \varphi (g)
\eea
The kernel of $\varphi$, is the set ${\rm ker}(\varphi)=\{f\in M:\varphi(f) = 0\}$ and the image of $\varphi$ is the set ${\rm im}(\varphi)=\{g\in N:{\rm there\,\, exists\,\,} f\in M {\rm\,\,with\,\,}\varphi (f) = g\}$. ${\rm ker}(\varphi)$ is a submodule of $M$ and ${\rm im}(\varphi)$ is a submodule of $N$.

Hilbert's syzygy theorem proves that every graded module $M$ over a graded ring $\cR$ has a graded resolution, meaning there exists an exact sequence
\bea
\begin{tikzcd}
  0 \arrow[r," "] & {\cal V}^{(k)}  \arrow[r, "\varphi_{k\,k-1}"] & {\cal V}^{(k-1)} \arrow[r, "\varphi_{k-1\,k-2}"] &\cdots \arrow[r, "\varphi_{21}"] &{\cal V}^{(1)} \arrow[r, "\varphi_{10}"] & {\cal V}^{(0)}  \arrow[r,"\varphi_0"] & M \arrow[r," "] & 0
\end{tikzcd}\label{HSexact}
\eea
where the ${\cal V}^{(i)}$ are free graded modules, and the arrows are graded linear maps of degree zero. Since not all modules $M$ have bases one needs both a generating set and the set of all relations satisfied by the image of the generators of ${\cal V}^{(0)}$. These relations are needed to decide if two elements expressed in terms of the generators are equal or not. The complete set of relations determine the kernel of $\varphi_0$ which is also equal to the image of $\varphi_{10}$ since the sequence (\ref{HSexact}) is exact. There is a correspondence between these relations and the generators of the module ${\cal V}^{(1)}$. The module ${\cal V}^{(1)}$ is called the first syzygy module and these generators are called the first syzygies. There are also relations between the images of these syzygies, which determines the kernel of $\varphi_{10}$, also equal to the image of $\varphi_{21}$. The syzgies corresponding to these relations generate the second syzygy module. Hilbert's syzygy theorem asserts that, if one continues in this way, starting with a module over a polynomial ring in $n$ indeterminates over a field, one eventually finds a zero module of relations, after at most $n$ steps\cite{syzygy}. The modules generated in this way, as well as the $\cR$-module homomorphisms are efficiently encoded in an exact sequence, called the resolution of $M$.

 In the next section we will see that we can recover an irreducible representation of a Lie algebra $\mg$  by imposing relations on a ring. The ring that is relevant here is $U(\mg_c^+)$, where we use the Cartan decomposition $\mg=\mg^+\oplus\mg^0\oplus\mg^-$, with $\mg^{0}$ the Cartan subalgebra, $\mg^{+}$ the raising operators and $\mg^-$ the lowering operators, and recall that $\mg^+_c$ is a commuting subalgebra of $\mg^+$. $U(\mg_c^+)$ is the universal enveloping algebra of $\mg_c^+$, which is a polynomial ring since $\mg_c^+$ is commutative.   The irreducible representation of the Lie algebra is itself a module $M$ of $U( \mg_c^+)$. In the light of the discussion above, the precise relation between the polynomial ring $\cR = U ( \mg_c^+) $ and the module $M$ is provided by the graded resolution above, with ${\cal V}^{(0)}$ being the ring $\cR$ and ${\cal V}^{(1)}$ the module built on the constraints needed to recover the Lie algebra irreducible representation from $M$.

A natural counting function associated with a graded module $M$ is the Hilbert function. For a finitely generated graded module $M$ over $R=k[x_0,\cdots, x_n]$, the Hilbert function ${\rm HM}(n)$ is
\bea
{\rm HM}(n) = {\rm dim}_k M_n
\eea
where ${\rm dim}_k M_n$ is the dimension of the vector space over the field $k$ of $M_n$,  the degree $n$ homogeneous subspace of $M$. As an example, if we take the graded module as $\cR=k[x_0,\cdots, x_q]$ itself, then $\cR_n$ is the vector space of homogeneous polynomials of degree $n$ in $q+1$ variables so that for $n\ge 0$ we have
\bea
{\rm HR}(t) = {\rm dim}_k R_n =\binom{n+q}{q}
\eea
The Hilbert function is used to define the Hilbert series (also called the Hilbert-Poincare series)
\bea
{\rm H}_M(t)=\sum_{n=0}^{\infty }{\rm HM}(n)~ t^{n}
\eea
Our goal is to use polynomial rings to study Lie algebra representations. The Hilbert series is then intrinsically interesting because it has a close relationship to the character of an irreducible representation. To see this, consider the character of group element $e^{\mu h}$ in irrep $M$, denoted $\chi_M(e^{\mu h})$ where $h\in\mg^0$. Use a basis given by the eigenbasis of $h$. Further, normalize $h$ so that its eigenvalues are integers. We then have
\bea
\chi_M(e^{\mu h})={\rm H}_M(e^\mu)
\eea

As we have mentioned above, the resolution of an irreducible representation viewed as a module of $U(\mg^+_c)$, is closely related to the character for that representation. As an illustration of this, we will explain how the exact sequence (\ref{HSexact}) can be used to obtain the character. Towards this end, recall that a short exact sequence is an exact sequence of the form
\bea\label{ShExSe} 
\begin{tikzcd}
  0 \arrow[r," "] & A  \arrow[r, "f"] & B \arrow[r, "g"] &C \arrow[r," "] & 0
\end{tikzcd}
\eea
$A$ is a subobject of $B$ with $f$ embedding $A$ into $B$. $C$ is the corresponding quotient $B/A$ with $g$ inducing an isomorphism
\bea
 C\cong B/{\rm im} (f)
\eea
%
%
%
Useful references are \cite{exactsequence,quotientspace}. A particularly useful character is the trace over a group element of the form $e^{\sum_i\mu_i H_i}$ with $H_i$ in the Cartan and $\mu_i$ are parameters of the character. The short exact sequence (\ref{ShExSe}) implies that the sum of traces, with alternating sign, vanishes
\bea
0=\Tr_A (e^{\sum_i\mu_i H_i})-\Tr_B (e^{\sum_i\mu_i H_i})+\Tr_C (e^{\sum_i\mu_i H_i})\label{dsum}
\eea
An exact sequence of modules
\bea
\begin{tikzcd}
  0 \arrow[r,"f_1"] & V_1  \arrow[r, "f_2"] & V_2 \arrow[r, "f_3"] &V_3 \cdots & V_{r-1}  \arrow[r,"f_{r}"] & V_r \arrow[r,"f_{r+1}"] & 0
\end{tikzcd}\label{lexact}
\eea
can be split into the short exact sequences
\bea
&&
\begin{tikzcd}
\qquad \qquad   0 \arrow[r," "] 
  & V_1 \arrow[r, " "] &K_2 \arrow[r," "] & 0
\end{tikzcd}\cr
&&
\begin{tikzcd}
  0 \arrow[r," "] & K_2  \arrow[r, " "] & V_2 \arrow[r, " "] &K_3 \arrow[r," "] & 0
\end{tikzcd}\cr
&&\qquad\qquad\vdots\qquad\qquad\vdots\qquad\qquad\vdots\cr
&&
\begin{tikzcd}
  0 \arrow[r," "] & K_{r-1}  \arrow[r, " "] & V_{r-1} \arrow[r, " "] &K_{r} \arrow[r," "] & 0
\end{tikzcd}\label{sexact}
\eea
where $K_i={\rm im}(f_i)$ for every $i$ and we have $K_r={\rm im}(f_r)=V_r$. The sequence (\ref{lexact}) is a long exact sequence if and only if (\ref{sexact}) are all short exact sequences \cite{exactsequence}. From (\ref{dsum}) it is now obvious that
\bea
\sum_{i=1}^r (-1)^i \Tr_{V_i}(e^{\sum_i\mu_i H_i}) = 0
\label{UsedbySage}
\eea
which, when applied to (\ref{su3sequence}) gives a formula for the character, after taking $\Tr_{V^{\rm phys}}(e^{\sum_i\mu_i H_i})$ to the right hand side.

To summarize the above discussion, two key ingredients, resolutions and Hilbert series, play a central role in understanding Lie algebra representations using polynomial rings and their modules. Fortunately, both are easily computed using the free open-source mathematics software system SageMath. As an example, consider the polynomial ring $\mathbb{R}[x,y]$ and set to zero everything in the ideal generated by $f_1$, $f_2$, $f_3$ and $f_4$ with
\bea
f_1=y^3\qquad f_2=xy^2\qquad f_3=x^2y\qquad f_4=x^3\label{firstideal}
\eea
This sets all polynomials of degree 3 or higher to zero, so that all that remains is
\bea
1,\,\,x,\,\,y,\,\,x^2,\,\,y^2,\,\,xy\label{polys}
\eea
To compute the Hilbert series we evaluate the code\footnote{This code can be pasted directly into the SageMathCell available at https://sagecell.sagemath.org/.}
\begin{verbatim}
names = '(x,y)'
Igens = ('x^3','x^2*y','x*y^2','y^3')
R = singular.ring(0,names,'dp')
I = singular.ideal(Igens)
from sage.rings.polynomial.hilbert import hilbert_poincare_series
hilbert_poincare_series(I)
\end{verbatim}
The resulting Hilbert polynomials is $3t^2 + 2t + 1$ which nicely matches (\ref{polys}). Sage evaluates this Hilbert series using the resolution (\ref{UsedbySage}) based on the ideal (\ref{firstideal}). To obtain the complete resolution we evaluate the code
\begin{verbatim}
names = '(x,y)'
Igens = ('x^3','x^2*y','x*y^2','y^3')
R = singular.ring(0,names,'dp')
I = singular.ideal(Igens)
singular.res(I,0)
\end{verbatim}
The resulting resolution is
\bea 
\begin{tikzcd}
  0 \arrow[r," "] & {\cal V}^{(2)}  \arrow[r, "\varphi_{21}"] & {\cal V}^{(1)} \arrow[r, "\varphi_{10}"] & {\cal V}^{(0)}  \arrow[r,"\varphi_{0*}"] &V \arrow[r," "] & 0
\end{tikzcd}
\label{su3sequence}
\eea
where ${\cal V}^{(0)}$ is the ring $\mathbb{R}[x,y]$, ${\cal V}^{(1)}=\langle f_1,f_2,f_3\rangle$ and ${\cal V}^{(2)}=\langle g_1,g_2,g_3\rangle$ with
\bea
g_1=-xf_1+yf_2\qquad g_2=-xf_2+yf_3\qquad g_3=-xf_3+yf_4
\eea
Here $V$ is the ring $\mathbb{R}[x,y]$ after setting the ideal $\langle f_1,f_2,f_3\rangle$ to zero. Notice that two syzygy modules ${\cal V}^{(1)}$ and ${\cal V}^{(2)}$ appear, in agreement with Hilbert's Syzygy Theorem.

\subsection{Resolutions of Lie Algebra Representations by Ring Modules}\label{RingsfromLieAlgebras}

Rings naturally arise from Lie algebra representations. A simple example illustrating the point is the fundamental representation of su(3). Introduce column vectors $(e_a)_i=\delta_{ai}$, $a,i=1,2,3$ as well as 3$\times$3 matrices $(E_{ab})_{ij}=\delta_{ai}\delta_{bj}$, $a,b,i,j=1,2,3$. In the Cartan decomposition of $\mg=$su(3) we have
\bea 
\mg = \mg^{ + }  \oplus \mg^{0} \oplus \mg^{ - } 
\eea
where $ \mg^{0}$ is the Cartan subalgebra spanned by $H_1=E_{11}-E_{22}$ and $H_2=E_{11}-E_{33}$. The Lie subalgebra $\mg^{+}$ is spanned by the raising operators  of su(3), which are $E_{32},E_{21},E_{31}$, while the Lie subalgebra $\mg^{-}$ is spanned by the lowering operators $E_{23},E_{12},E_{13}$. The lowest weight vector for the fundamental representation $e_1$, is annihilated by $E_{32}$. Applying $E_{21},E_{31}$ we generate
\bea
e_2=E_{21}e_1\qquad e_3=E_{31}e_1\label{gentwostates}
\eea
Thus, starting from $e_1$ and the commuting subalgebra $\mg_c^+=\{E_{21},E_{31}\}$, we generate the complete irreducible representation. Since we use a commuting subalgebra $\mg_c$, it is natural to expect that the commutative polynomial ring\footnote{The choice of the Cartan subalgebra made above is not unique. This choice ensures that the eigenvalues of elements of the Cartan are simply related to the degree in $x_{21}$ and the degree in $x_{31}$.} ${\cal V}^{(0)}=\mathbb{C}[x_{21},x_{31}]$ provides a useful starting point to study this irrep. This expectation is bolstered by the observation that ${\cal V}^{(0)}$ admits a representation of U(su(3)). The su(3) generators are constructed from the $E_{ab}$ which are realized as differential operators, as follows
\bea
\hat E_{11}&=&-x_{21}{\partial\over\partial x_{21}}-x_{31}{\partial\over\partial x_{31}}+1\qquad
\hat E_{22}\,\,\,=\,\,\,x_{21}{\partial\over\partial x_{21}}\qquad
\hat E_{33}\,\,\,=\,\,\,x_{31}{\partial\over\partial x_{31}}\cr\cr
\hat E_{23}&=&x_{21}{\partial\over\partial x_{31}}\qquad\qquad
\hat E_{12}\,\,\,=\,\,\,-{1\over 2}x_{21}{\partial\over\partial x_{21}}{\partial\over\partial x_{21}}
-x_{31}{\partial\over\partial x_{31}}{\partial\over\partial x_{21}}+{\partial\over\partial x_{21}}\cr\cr
\hat E_{32}&=&x_{31}{\partial\over\partial x_{21}}\qquad\qquad
\hat E_{13}\,\,\,=\,\,\,-{1\over 2}x_{31}{\partial\over\partial x_{31}}{\partial\over\partial x_{31}}
-x_{21}{\partial\over\partial x_{21}}{\partial\over\partial x_{31}}+{\partial\over\partial x_{31}}\cr
&&\label{su3gens}
\eea
and $\hat E_{21}=x_{21}$, $\hat E_{31}=x_{31}$. These differential operators are hatted to distinguish them from the 3$\times$3 matrices introduced above. The Lie algebra generators $E_{21}$ and $E_{31}$ square to zero and their product is zero
\bea
E_{21}E_{21}=0\qquad E_{31}E_{31}=0\qquad E_{21}E_{31}=0
\eea
as is easily checked using the explicit matrix representation given at the start of this subsection. Consequently to obtain su(3) from ${\cal V}^{(0)}$ we must impose the relations
\bea
f_1=x_{21}^2=0\qquad f_2=x_{31}^2=0\qquad f_3=x_{21}x_{31}=0\label{deff1f2f3}
\eea
To confirm that this does indeed produce the fundamental representation of su(3), we compute the Poincare-Hilbert polynomial using Sage
\begin{verbatim}
names = '(x21,x31)'
Igens = ('x21*x21','x21*x31','x31*x31')
R = singular.ring(0,names,'dp')
I = singular.ideal(Igens)
from sage.rings.polynomial.hilbert import hilbert_poincare_series
hilbert_poincare_series(I)
\end{verbatim}
The result is
\bea
P(t) = 2t + 1
\eea
We immediately read off that there are 3 states in the irrep, with one of them ($e_1$) a constant polynomial and two of them ($e_2$ and $e_3$) degree one polynomials in agreement with (\ref{gentwostates}). Computing the associated resolution using Sage
\begin{verbatim}
names = '(x21,x31)'
Igens = ('x21*x21','x21*x31','x31*x31')
R = singular.ring(0,names,'dp')
I = singular.ideal(Igens)
singular.res(I,0)
\end{verbatim}
we find the following exact sequence
\bea
\begin{tikzcd}
  0 \arrow[r," "] & {\cal V}^{(2)}  \arrow[r, "\varphi_{21}"] & {\cal V}^{(1)} \arrow[r, "\varphi_{10}"] & {\cal V}^{(0)}  \arrow[r,"\varphi_{0*}"] & V^{\rm phys} \arrow[r," "] & 0
\end{tikzcd}
\label{su3sequence}
\eea
The module ${\cal V}^{(1)}=\langle f_1,f_2,f_3\rangle$ with $f_1,f_2$ and $f_3$ defined in (\ref{deff1f2f3}) and ${\cal V}^{(2)}=\langle g_1,g_2\rangle$ where
\bea
g_1=-x_{21}f_2+x_{31}f_3\qquad g_2=-x_{21}f_3+x_{31}f_1\label{2ndfundsyz}
\eea
$V^{\rm phys}$ is the fundamental representation of su(3). Thus, starting with the polynomial ring ${\cal V}^{(0)}$ we have constructed a resolution of $V^{\rm phys}$ in terms of modules of the ring. The maps appearing in the sequence are defined as follows
\bea
\varphi_{0*}(1)=e_1\qquad\varphi_{0*}(x_{21})=e_2\qquad\varphi_{0*}(x_{31})=e_3
\eea
and $\varphi_{0*}(m)=0$ with $m$ any monomial of degree greater than 1. Since module homomorphism are ${\cal V}^{(0)}$-linear this completes the definition of $\varphi_{0*}$. Similarly, we define $\varphi_{10}$ by specifying
\bea
\varphi_{10}(f_1)=x_{21}^2\qquad\varphi_{10}(f_2)=x_{31}^2\qquad
\varphi_{10}(f_3)=x_{21}x_{31}
\eea
Since everything in the image of $\varphi_{10}$ has degree greater than 1, it is clear that ${\rm im}(\varphi_{10})={\rm ker}(\varphi_{0*})$, as it must be for an exact sequence. $\varphi_{21}$ is defined by
\bea
\varphi_{21}(g_1)=-x_{21}f_2+x_{31}f_3\qquad
\varphi_{21}(g_2)=-x_{21}f_3+x_{31}f_1
\eea
Noting that
\bea
\varphi_{10}(-x_{21}f_2+x_{31}f_3)&=&
-x_{21}\varphi_{10}(f_2)+x_{31}\varphi_{10}(f_3)
=-x_{21}x_{31}^2+x_{31}^2x_{21}=0\cr\cr
\varphi_{10}(-x_{21}f_3+x_{31}f_1)&=&
-x_{21}\varphi_{10}(f_3)+x_{31}\varphi_{10}(f_1)
=-x_{21}^2x_{31}+x_{31}x_{21}^2=0
\eea
It is clear that ${\rm im}(\varphi_{21})={\rm ker}(\varphi_{10})$ which confirms that (\ref{su3sequence}) is indeed an exact sequence.

The exact sequence (\ref{su3sequence}) is not su(3) equivariant. To see this, note that $\varphi_{10}$  embeds ${\cal V}^{(1)}$ in ${\cal V}^{(0)}$. su(3) equivariance requires that the subspace of ${\cal V}^{(0)}$ given by the image of ${\cal V}^{(1)}$ under $\varphi_{10}$ is an invariant subspace of U(su(3)). It is easy to see that this is not the case, using the explicit generators in (\ref{su3gens})
\bea
\hat E_{12}\varphi_{10}(f_1)=-x_{21}
\eea
Since $x_{21}$ is degree 1 it is not in the image of $\varphi_{10}$. The exact sequence (\ref{su3sequence}) is equivariant under the smaller group, generated by the Cartan of su(3) ($\hat H_1,\hat H_2$) and the elements of $\mg_c$. To demonstrate equivariance, we start by assigning the following charges to module states
\bea
\hat H_1(x_{21}^m x_{31}^n)&=&(1-2m-n)x_{21}^m x_{31}^n\qquad\qquad
\hat H_2(x_{21}^m x_{31}^n)\,\,=\,\,(1-m-2n)x_{21}^m x_{31}^n\cr\cr
\hat H_1(x_{21}^m x_{31}^nf_1)&=&-(3+2m+n)x_{21}^m x_{31}^nf_1\qquad\qquad
\hat H_2(x_{21}^m x_{31}^nf_1)\,\,=\,\,-(1+m+2n)x_{21}^m x_{31}^nf_1\cr\cr
\hat H_1(x_{21}^m x_{31}^nf_2)&=&-(1+2m+n)x_{21}^m x_{31}^nf_2\qquad\qquad
\hat H_2(x_{21}^m x_{31}^nf_2)\,\,=\,\,-(3+m+2n)x_{21}^m x_{31}^nf_2\cr\cr
\hat H_1(x_{21}^m x_{31}^nf_3)&=&-(2+2m+n)x_{21}^m x_{31}^nf_3\qquad\qquad
\hat H_2(x_{21}^m x_{31}^nf_3)\,\,=\,\,-(2+m+2n)x_{21}^m x_{31}^nf_3\cr\cr
\hat H_1(x_{21}^m x_{31}^ng_1)&=&-(3+2m+n)x_{21}^m x_{31}^ng_1\qquad\qquad
\hat H_2(x_{21}^m x_{31}^ng_1)\,\,=\,\,-(4+n+2m)x_{21}^m x_{31}^ng_1\cr\cr
\hat H_1(x_{21}^m x_{31}^ng_2)&=&-(4+2m+n)x_{21}^m x_{31}^ng_2\qquad\qquad
\hat H_2(x_{21}^m x_{31}^ng_2)\,\,=\,\,-(3+m+2n)x_{21}^m x_{31}^ng_2\cr
&&
\eea
To see how equivariance is established consider, for example $p=x_{21}^nx_{31}^mf_1\in{\cal V}^{(1)}$. Using $\hat E_{21}p=x_{21}p$ and
\bea
\varphi_{10}(p)\,\,=\,\,x_{21}^{n+2}x_{31}^m\in{\cal V}^{(0)}
\eea
we have
\bea
\varphi_{10}(\hat E_{21}p)&=&\varphi_{10}(x_{21}^{n+1}x_{31}^mf_1)
=x_{21}^{n+3}x_{31}^m\cr\cr
&=&\hat E_{21}x_{21}^{n+2}x_{31}^m
=\hat E_{21}\varphi_{10}(p)
\eea
Very similar arguments establish that
\bea
\varphi_{10}(\hat E_{31}p)&=&\hat E_{31}\varphi_{10}(p)\cr\cr
\varphi_{10}(\hat H_{1}p)&=&\hat H_{1}\varphi_{10}(p)\cr\cr
\varphi_{10}(\hat H_{2}p)&=&\hat H_{2}\varphi_{10}(p)
\eea
which proves the equivariance of $\varphi_{10} (\cdot)$. It is equally easy to check equivariance for $\varphi_{0*}$ and $\varphi_{21}$.

The algebra for the equivariance group is
\bea
[H_1 ,E_{21}]=-2E_{21}\qquad [H_2 ,E_{21}]=-E_{21}
\eea
\bea
[H_1 , E_{31}]=-E_{31}\qquad [H_2 ,E_{31}]=-2E_{31}
\eea
Introduce
\bea
A_1=E_{21}\qquad A_2={1\over 3}E_{11}-{2\over 3}E_{22}+{1\over 3}E_{33}
\eea
\bea
\tilde{A}_1=E_{31}\qquad \tilde{A}_2={1\over 3}E_{11}+{1\over 3}E_{22}-{2\over 3}E_{33}
\eea
which obey
\bea
[A_1,A_2]=A_1 \qquad [\tilde{A}_1,\tilde{A}_2]=\tilde{A}_1\qquad [A_i,\tilde{A}_j]=0
\eea
This is two copies of commuting copies of $L(2,0)$ (see \cite{Bowers} for the classification of three dimensional real Lie algebras) so that our equivariance group has Lie algebra $L(2,0)\oplus L(2,0)$.

In the above discussion we are setting the $\mg^0$ charges of the ground states of the higher depth modules to be different from the ones of the lowest depth module.  This is necessary if we are to have equivariance under $\mg^+_c\oplus \mg^0$. When we come to discuss the half BPS representation later, we will not have the luxury of freely choosing the $\mg^0$ charges, since their values are constrained by the BPS condition. Consequently, it might not be possible to include all of $\mg^0$ in the equivariance group in the half BPS case.

\subsection{Comparison with BGG Resolutions}\label{BGG}

As discussed above, our approach employing polynomial rings is inspired by the BGG resolution and properties of the half BPS representation in $\cN=4$ super Yang-Mills theory, so it is useful to now describe the BGG resolution in light of the examples we have just considered. The modules that appear in the exact sequence of the BGG resolution are all $U(\mg)$ lowest weight Verma modules \cite{HWM}. A lowest weight Verma module need not be irreducible, due to the appearance of null states. In this case, the representation is indecomposable and the structure of the null states which appear is reflected in the exact sequence of the resolution. The BGG resolution can be used to recover the Weyl character formula.

The weights of the lowest weight modules appearing in the BGG exact sequence are related to the weight of the representation being resolved by the action of the Weyl group. The Weyl group for su(3) is generated by two elements $\sigma_1$ and $\sigma_2$. Their action on the u(3) weights is as follows \cite{BDHO}
\bea
\sigma_1\cdot (l_1,l_2,l_3)=(l_2-1,l_1+1,l_3)\cr\cr
\sigma_2\cdot (l_1,l_2,l_3)=(l_1,l_3-1,l_2+1)
\eea
The other elements of the Weyl group are $\sigma_1\cdot\sigma_2$, $\sigma_2\cdot \sigma_1$ and  $\sigma_1\cdot\sigma_2\cdot\sigma_1$. The Weyl group for su(3) is isomorphic to the symmetric group $S_3$. Restricting to su(3) is achieved by imposing $l_1+l_2+l_3=0$. This is invariant under $S_3$ so the Weyl group maps su(3) representations into each other.

The fundamental representation $V^F$ of su(3) has lowest weight $l_F=(1,0,0)$.
Under the action of the Weyl group we generate the following additional weights
\bea
\sigma_1 \cdot l_F &=&(-1,2,0)\qquad\qquad \sigma_2 \cdot l_F\,\,=\,\,(1,-1,1)\cr\cr
\sigma_1\cdot\sigma_2 \cdot l_F&=&(-2,2,1)\qquad\qquad
\sigma_2\cdot \sigma_1 \cdot l_F\,\,=\,\,(-1,-1,3)\cr\cr
\sigma_1\cdot\sigma_2\cdot\sigma_1 \cdot l_F&=&(-2,0,3)\label{FWeights}
\eea
The BGG resolution is
\bea
\begin{tikzcd}
  0 \arrow[r," "] & \cV^{(3)}  \arrow[r, "\varphi_{32}"] & \cV^{(2)}\arrow[r, "\varphi_{21}"] & \cV^{(1)}\arrow[r,"\varphi_{10}"] &\cV^{(0)}\arrow[r,"\varphi_{0*}"] & V^F \arrow[r," "] & 0
\end{tikzcd}
\eea
\bea
\cV^{(0)}&=&{\cal V}_{l_F}\qquad \qquad\qquad
\cV^{(1)}\,\,\,=\,\,\,{\cal V}_{\sigma_2\cdot l_F}\oplus{\cal V}_{\sigma_1\cdot l_F}\cr\cr
\cV^{(2)}&=&{\cal V}_{\sigma_1\cdot\sigma_2\cdot l_F}\oplus {\cal V}_{\sigma_2\cdot\sigma_1\cdot l_F}\qquad
\cV^{(3)}\,\,\,=\,\,\,{\cal V}_{\sigma_1\cdot\sigma_2\cdot\sigma_2\cdot l_F}
\eea
where in the last equation above we have labelled each lowest weight module by the lowest weight. These are all modules over the ring $U(\mg_+)$, given by polynomials in $E_{21}$, $E_{31}$ and $E_{32}$. Note that this is not a commutative ring. The module ${\cal V}_{l_F}$ is constructed on the lowest weight state $|e_1\rangle$, which has the same charge under the Cartan as $e_1$ does. The Hilbert series of  ${\cal V}_{l_F}$ is
\bea
H_{{\cal V}_{l_F}}=H_{{\cal V}_{(1,0,0)}}=
{u_1\over (1-{u_2\over u_1})(1-{u_3\over u_1})(1-{u_3\over u_2})}
\eea
where we have used the variable $u_i$ to track the $E_{ii}$ charge. In the fundamental $V^F$ of su(3) we have the following identities
\bea
E_{21}E_{21}e_1=0\qquad
E_{32}e_1=0\qquad
E_{31}E_{31}e_1=0\qquad
E_{31}E_{21}e_1=0\label{constrat}
\eea
Since each of these appears in ${\cal V}_{l_F}$, they must lie in the kernel of $\varphi_{0*}:{\cal V}_{l_F}\to V^F$. Requiring that $E_{21}E_{21}e_1$ and $E_{32}e_1$ are in the kernel is enough since
\bea
0=E_{32}E_{21}E_{21}e_1=[E_{32},E_{21}]E_{21}e_1+E_{21}[E_{32},E_{21}]e_1
=-2E_{21}E_{31}e_1
\eea
and hence
\bea
0=E_{32}E_{21}E_{31}e_1=E_{31}E_{31}e_1=0
\eea
Consequently, to obtain the fundamental representation of su(3) we must quotient ${\cal V}_{l_F}$ by the direct sum of the lowest weight module built on the state $|E_{21}E_{21}e_1\rangle$ and the lowest weight module built on the state $|E_{32}e_1\rangle$. These states have weight $(-1,2,0)=\sigma_1\cdot l_F$ and $(1,-1,1)=\sigma_2\cdot l_F$, so that we need to quotient by ${\cal V}_{\sigma_1\cdot l_F}\oplus {\cal V}_{\sigma_2\cdot l_F}$. The corresponding Hilbert series are
\bea
H_{{\cal V}_{\sigma_1\cdot l_F}}=
{u_1^{-1}u_2^2 \over (1-{u_2\over u_1})(1-{u_3\over u_1})(1-{u_3\over u_2})}
\qquad
H_{{\cal V}_{\sigma_2\cdot l_F}}=
{u_1u_2^{-1}u_3 \over (1-{u_2\over u_1})(1-{u_3\over u_1})(1-{u_3\over u_2})}
\eea
The module homomorphism $\varphi_{10}$ is defined by
\bea
\varphi_{10}(|E_{21}E_{21}e_1\rangle)=E_{21}E_{21}|e_1\rangle\qquad
\varphi_{10}(|E_{32}e_1\rangle)=E_{32}|e_1\rangle
\eea
The modules ${\cal V}_{\sigma_1\cdot l_F}$ and ${\cal V}_{\sigma_2\cdot l_F}$ overlap. For example, $E_{32}E_{31}E_{31}|e_1\rangle$ is in the overlap since
\bea
\varphi_{10}(E_{32}E_{32}E_{32}|E_{21}E_{21}e_1\rangle)=E_{32}E_{31}E_{31}|e_1\rangle= \varphi_{10}(E_{31}E_{31}|E_{32}e_1\rangle)
\eea
The repeated states fills out the direct sum of a lowest weight module constructed on the state with weight $(-2,2,1)=\sigma_1\cdot\sigma_2\cdot l_F$ and a lowest weight module constructed on the state with weight $(-1,-1,3)=\sigma_2\cdot\sigma_1\cdot l_F$. The Hilbert series of these modules are
\bea
H_{{\cal V}_{\sigma_1\cdot\sigma_2\cdot l_F}}={u_1^{-2}u_2^2u_3\over (1-{u_2\over u_1})(1-{u_3\over u_1})(1-{u_3\over u_2})}\qquad
H_{{\cal V}_{\sigma_2\cdot\sigma_1\cdot l_F}}={u_1^{-1}u_2^{-1}u_3^3 \over (1-{u_2\over u_1})(1-{u_3\over u_1})(1-{u_3\over u_2})}
\eea
The modules ${\cal V}_{\sigma_1\cdot\sigma_2\cdot l_F}$ and ${\cal V}_{\sigma_2\cdot\sigma_1\cdot l_F}$ also overlap, with the overlap given by the lowest weight module constructed on the state with weight $(-2,0,3)=\sigma_1\cdot\sigma_2\cdot \sigma_1\cdot l_F$. The Hilbert series of this module is
\bea
H_{{\cal V}_{\sigma_1\cdot\sigma_2\cdot \sigma_1\cdot l_F}}={u_1^{-2}u_3^3 \over (1-{u_2\over u_1})(1-{u_3\over u_1})(1-{u_3\over u_2})}
\eea
Since this is a single module, there is no possible overlap, the kernel of $\varphi_{32}$ is empty and the exact sequence terminates.

The Weyl character formula is
\bea
\chi_{(l_1,l_2,l_3)}={N\over (u_1-u_2)(u_1-u_3)(u_2-u_3)}
\eea
where
\bea
N=\det\left[\begin{array}{ccc}
u_1^{l_1+2} &u_1^{l_2+1} &u_1^{l_3}\\
u_2^{l_1+2} &u_2^{l_2+1} &u_2^{l_3}\\
u_3^{l_1+2} &u_3^{l_2+1} &u_3^{l_3}
\end{array}\right]
\eea
Specializing to the fundamental of su(3), we have $l_1=1$, $l_2=0$ and $l_3=0$. The weights (\ref{FWeights}) make a visible appearance as the powers of $u_i$, after a simple rewriting
\bea
\chi_{V^F}&=&{u_1^3 u_2-u_2^3u_1-u_1^3u_3+u_2^3u_3+u_3^3u_1-u_3^3u_2
\over (u_1-u_2)(u_1-u_3)(u_2-u_3)}\cr\cr
&=&{u_1-u_1^{-1}u_2^2-u_1u_2^{-1}u_3+u_1^{-2}u_2^2u_3+u_1^{-1}u_2^{-1}u_3^3-u_1^{-2}u_3^3
\over (1-{u_2\over u_1})(1-{u_3\over u_1})(1-{u_3\over u_2})}\label{WeylChar}
\eea
Each term in the numerator corresponds to a unique lowest weight module and they have been combined to produce the character exactly as dictated by the BGG resolution.

To prove that the BGG resolution is an exact sequence, we need to prove (i) that ${\rm Im}(\varphi_{i+1\,i})={\rm Ker}(\varphi_{i\,i-1})$ for $i=0,1,2$ and (ii) that the above maps are equivariant. The maps appearing have the correct equivariance properties as a consequence of the fact that lowest weight modules are invariant subspaces of U(su(3)). The reader interested in further details can consult \cite{BGG,BGGCat}. This completes our description of the BGG resolution.

\subsection{Super-rings and resolutions}\label{supering}

Our ultimate goal is a description of the fundamental field representation of ${\cal N}=4$ super Yang-Mills theory. The relevant ring necessarily includes Grassmann variables. In this section we introduce a simple example of a ring of Grassman variables. Using this example, we explain how both the Hilbert series and resolutions of a super-ring can be evaluated using Sage.

An instructive toy model is provided by the ring consisting of three anticommuting variables, $Q_1$, $Q_2$ and $Q_3$, together with the constraint
\bea
Q_1 Q_2 =0\label{GrssmanIdl}
\eea
The ``state space'' $V^Q$ corresponds to the following polynomials
\bea
1,Q_1,Q_2,Q_3,Q_1Q_3,Q_2Q_3
\eea
so that we have 3 bosons and 3 fermions. The corresponding Hilbert series is
\bea
H_{V^Q}(t)=1+3t+2t^2\label{Gresltn}
\eea
Sage is able to evaluate the Hilbert series using the resolution (\ref{UsedbySage}) based on the ideal (\ref{GrssmanIdl}). As a test of the resolution, the answer from Sage should agree with (\ref{Gresltn}). Since we have Grassmann valued variables, we are considering a super commutative ring. Sage is able to treat super commutative rings by calling functions from the superCommutative library available in singular. To obtain the Hilbert series, we execute the following code: 
\begin{verbatim}
names = '(q1,q2,q3)'
Igens = ('q1*q2')
R = singular.ring(0,names,'dp')
A = singular.superCommutative(1,3)
A.set_ring()
I = singular.ideal(Igens)
from sage.rings.polynomial.hilbert import hilbert_poincare_series
hilbert_poincare_series(I)
\end{verbatim}
The line 'A = singular.superCommutative(1,3)' declares that variables 1 to 3 inside ``names'' are Grassmann valued. The result from Sage is in perfect agreement with (\ref{Gresltn}), providing support for the resolution constructed from the ideal (\ref{GrssmanIdl}). To construct the resolution itself, we execute the following code:
\begin{verbatim}
names = '(q1,q2,q3)'
Igens = ('q1*q2')
R = singular.ring(0,names,'dp')
A = singular.superCommutative(1,3)
A.set_ring()
I = singular.ideal(Igens)
singular.res(I,0)
\end{verbatim}
The result gives us the resolution
\bea
\begin{tikzcd}
  0 \arrow[r," "] & {\cal V}^{(3)}  \arrow[r, "\varphi_{32}"] & {\cal V}^{(2)}  \arrow[r, "\varphi_{21}"] & {\cal V}^{(1)}\arrow[r, "\varphi_{10}"] & {\cal V}^{(0)}\arrow[r,"\varphi_{0*}"] & V^Q \arrow[r," "] & 0
\end{tikzcd}
\eea
where
\bea
\cV^{(1)}= {\cal V}_{\langle f\rangle}\qquad
\cV^{(2)}={\cal V}_{\langle g_1,g_2\rangle}\qquad
\cV^{(3)}={\cal V}_{\langle h_1,h_2,h_3\rangle}
\eea
and
\bea
f=Q_1Q_2
\eea
\bea
g_1=Q_2f\qquad g_2=Q_1 f
\eea
\bea
h_1=Q_2 g_1\qquad h_2=Q_1 g_1 +Q_2 g_2\qquad h_3=Q_1 g_2
\eea
In the next section we will develop examples in which the ring has both bosonic and fermionic generators. 

\section{psu$(2,2|4)$ symmetry of SYM and super-polynomial ring $\cR (8|8) $}\label{symmetryandring}

In this section we introduce the generators of $\mg=$psu$(2,2|4)$ and their commutation relations. We describe lowest weight primary states and then explain how  Verma modules of
$\mg=$psu$(2,2|4)$   are constructed by acting on these states with the enveloping algebra $U ( \mg^+)$ generated by  a subspace $ \mg^+ \subset \mg $.  Specializing to the half-BPS lowest weight primary states, we show that the subspace $ \mg_c^+ \subset \mg^+ $ which does   not annihilate the half BPS state is a commutative subspace. This motivates the definition of a super-polynomial ring $U(\mg_c^+)$, with eight bosonic and eight fermionic generators. We will also use the notation $ \mathcal{R}(8|8) = U ( \mg_c^+ ) $. Acting on the half-BPS state with $U ( \mg_c^+ ) $ we generate a module of the super-ring which we refer to as $V_{\cR}^{(0)}$. The states in $V_{\cR}^{(0)}$ can be organised according to  eigenvalues of the generators of $\mg^0$ which include the scaling dimension $\Delta$. For $ \Delta =1$, the module  $V_{\cR}^{(0)}$ is related to the quantum states in the fundamental field representation of U$(1) $ $ \cN =4$ super Yang-Mills theory.  We  explain how   the elementary fields of $ \cN =4$ super Yang-Mills theory (and their derivatives)  correspond  to low-dimension states in $V_{\cR}^{(0)}$. There are redundancies in the map  from $V_{\cR}^{(0)}$ to the space of elementary fields, some of which are due to the equations of motion and others are due to the fact that different elements of $U ( \mg_c^+ ) $ acting on the half-BPS state can give rise to the same state in the half-BPS representation. The precise description of these redundancies is given by resolution of the fundamental field representation $V_{\rm{phys}}^{\rm BPS}$  in terms of modules of $ \mathcal{R}(8|8) = U ( \mg_c^+ ) $. This resolution is described in the next section.

\subsection{A super-ring from half-BPS representations}\label{sec:SYMsuperring} 

The Lie algebra of the conformal group so(4,2)=su(2,2) consists of translation generators $P_{\alpha\dot{\alpha}}$, special conformal generators $K^{\dot{\alpha}\alpha}$, dilatations $D$ and su(2)$\otimes$su(2) spin generators $M_\alpha{}^\beta$ and $\bar{M}^{\dot\alpha}{}_{\dot\beta}$. We are using the bispinor notation, reviewed in Appendix \ref{conventions}, for spacetime tensors.  The superconformal extension psu$(2,2|4)$ includes the supercharges $Q^i{}_\alpha$, $\bar{Q}_{i\dot{\alpha}}$, their conformal partners $S_i{}^\alpha$ and $\bar{S}^{i\dot{\alpha}}$ and the su(4) R-symmetry generators $R^i{}_j$. Here $\alpha,\dot{\alpha}=1,2$ and $i=1,\dots, {\cal N}$ with ${\cal N}=4$. The Lie algebra of these generators is given in Appendix \ref{psu224}.

Let us call $\mathfrak{g} $ the Lie algebra psu$(2,2|4)$. We will use the Cartan decomposition
\bea 
\mg = \mg^{ + }  \oplus \mg^{0} \oplus \mg^{ - } 
\eea
where $ \mg^{0}$ is the Cartan subalgebra of psu$(2,2|4)$. 
The Lie subalgebra $\mg^{+}$ is spanned by the raising operators  of su(4), su(2,2) and the supercharges
$Q^i{}_\alpha$, $\bar{Q}_{j\dot{\beta}}$. 
The Lie subalgebra $\mg^{-}$ is spanned by the lowering operators  of su(4), su(2,2) and the conformal
supercharges $S_j{}^\beta$, $\bar{S}^{i\dot{\alpha}}$.

A generic lowest weight primary state $|{\cal O}\rangle_{\rm l.w.}\equiv |\Delta; k,p,q;j,\bar{j}\rangle_{\rm l.w.}$
for this superalgebra has conformal dimension $\Delta$, belongs to the 
su(2)$\otimes$su(2) representation $(j,\bar{j})$ and the su(4) representation with Dynkin labels $[k, p, q]$. 
It is annihilated by the lowering operators $\mg^{-}$: by the special conformal generator and the conformal supercharges
\bea
K^{\dot{\alpha}\alpha}|{\cal O}\rangle_{\rm l.w.}&=&0\qquad
S_j{}^\alpha|{\cal O}\rangle_{\rm l.w.}\,\,\,=\,\,\,0\qquad
\bar{S}^{i\dot{\alpha}}|{\cal O}\rangle_{\rm l.w.}\,\,\,=\,\,\,0
\eea
by the su(2)$\otimes$su(2) lowering operators and by the su(4) lowering operators
\bea
R^i{}_j|{\cal O}\rangle_{\rm l.w.}&=&0\qquad j<i
\eea
The action of the Cartan subalgebra $\mg^{0}$ is
\bea
D|{\cal O}\rangle_{\rm l.w.}&=&\Delta |{\cal O}\rangle_{\rm l.w.}\qquad
H_1|{\cal O}\rangle_{\rm l.w.}\,\,\,=\,\,\,k|{\cal O}\rangle_{\rm l.w.}\qquad
H_2|{\cal O}\rangle_{\rm l.w.}\,\,\,=\,\,\,p|{\cal O}\rangle_{\rm l.w.}\cr\cr
H_3|{\cal O}\rangle_{\rm l.w.}&=&q|{\cal O}\rangle_{\rm l.w.}\qquad
J_3|{\cal O}\rangle_{\rm l.w.}\,\,\,=\,\,\,j|{\cal O}\rangle_{\rm l.w.}\qquad
\bar{J}_3|{\cal O}\rangle_{\rm l.w.}\,\,\,=\,\,\,\bar{j}|{\cal O}\rangle_{\rm l.w.}
\eea
where $J_3=M_1{}^1$, $\bar{J}_3=\bar{M}^1{}_1$ and the $H_i$ are the Cartan generators for su(4), given by
\bea
H_i=R^i{}_i-R^{i+1}{}_{i+1}\qquad i=1,2,3
\eea
in terms of the su(4) generators introduced above.
By acting on the lowest weight primary state with the raising operators 
$\mg^+=\{P_{\alpha\dot{\alpha}},Q^i{}_\alpha,\bar{Q}_{i\dot{\beta}},J_+,\bar{J}_+,R^i{}_j\}$ with $j>i$, $J_+=M_1{}^2$ and $\bar{J}_+=\bar{M}^1{}_2$, we fill out the Verma module constructed on this lowest weight primary. We write this Verma module as $U(\mg^+)|{\cal O}\rangle_{\rm l.w.}$ where $U(\mg^+)$ is the universal enveloping  algebra of $\mg^+$.

Special lowest weight primary states corresponding to short (BPS) representations are also annihilated by certain
supercharges. Our focus in what follows, is on the half-BPS representation of ${\cal N}=4$ super Yang-Mills theory.
This is the shortest non-trivial multiplet possible and hence has a lowest weight  preserving the maximal amount of  supersymmetry. Such states correspond, under the operator-state correspondence, to multi-traces constructed from one complex matrix $Z$. The relevant BPS lowest weight primary state $|\cO_{BPS}\rangle_{\rm l.w.}$ is annihilated by the elements of $\mg^-$, and in addition obeys \cite{BDHO}
\bea
Q^i{}_\alpha| \cO_{BPS} \rangle_{\rm l.w.}&=&0\qquad
\bar{Q}_{i+2\,\dot{\alpha}}| \cO_{ BPS} \rangle_{\rm l.w.}\,\,\,=\,\,\,0
\eea
where $i,\alpha,\dot{\alpha}=1,2$, as well as
\bea
R^4{}_3| \cO_{BPS} \rangle_{\rm l.w.}&=&0\qquad
R^2{}_{1}| \cO_{ BPS} \rangle_{\rm l.w.}\,\,\,=\,\,\,0
\eea
Consequently, the set of generators that does not annihilate this state is denoted $\cS_{ + }^c $
\bea
\{P_{\alpha\dot\alpha}, \bar{Q}_{i\dot\alpha}, Q^{i+2}{}_\alpha,R^3{}_1, R^3{}_2, R^4{}_1, R^4{}_2\}
\equiv \cS^{ + }_c 
\label{commsubset} 
\eea
where $i,\alpha,\dot{\alpha}=1,2$.  Let us define $\mg_c^+$ to be space  of complex linear combinations of the generators in $\cS_{ +}^c$ : 
\bea 
\mg_c^+ = \hbox{ Span}_{ \mC } \{ \cS^{ +}_c \} 
\label{ingcplus}
\eea
  It is easily verified using the equations in Appendix \ref{psu224}  that $\mg_c^+$, which is a subspace of $\mg$ , is a graded commutative algebra:  the fermionic generators anti-commute with each other, while the bosonic generators commute with each other and with the fermions. $\mg_c^+$ is thus a graded Lie sub-algebra
of the graded Lie algebra $\mg^+$, which is itself a graded Lie sub-algebra of $\mg = \psu(2,2|4)$ : 
\bea 
\mg_c^+   \subset \mg^+ \subset \mg 
\eea
We also have a corresponding embedding of the universal enveloping algebras. Following the standard construction of enveloping algebras, we define $ U ( \mg ) $ as the graded associative algebra with generators given by those of $\mg$ and relations given by replacing graded Lie brackets of $\mg$ with graded commutators in $ U (\mg)$.  $U ( \mg^+ )$ and $U ( \mg_c^+) $ are similarly defined using the generators of $\mg^+$ and $\mg^+_c$ respectively. We also have 
\bea 
U ( \mg_c^+ )    \subset U ( \mg^+)  \subset U ( \mg )  
\eea

The conditions for short and semi-short reps for ${\cal N}=4$ superconformal symmetry given in \cite{BDHO} are given in terms of the fraction $=(t,\bar{t})$ of $(Q,\bar{Q})$ supersymmetries annihilating the superconformal primary, where $t,\bar{t}\in\{{1\over 2},{1\over 4},{1\over 8}\}$. This property that all the non-annihilating generators commute with each other is a special property of the half-BPS state\cite{BDHO}, which has $t=\bar{t}={1\over 2}$. There are eight  bosonic generators  and eight  fermionic generators in $\mg_c^+$. Thus it is also natural to identify  $ \mg_c^+ \equiv  \mR ( 8 |8)$. Let us call ${ \cal R}(8|8)$ the graded polynomial ring generated by the elements of $\cS^+_c$. We may write $ U ( \mg_c^+ ) \equiv \mC  [ \cS^+_c ] $. We summarise these observations as 

{\vskip 0.25cm}

{\bf Lemma 1: } The enveloping algebra $ U  ( \mg_c^+  ) \equiv \mC ( \cS^+_c )$ is the super-ring of polynomials with complex coefficients  generated by $\cS^+_c$.

\boxedeq{eq:thm1}{  U (\mg^+_c)=\mathbb{C}[\cS^+_c] = \cR ( 8 |8 ) = U ( \mR ( 8 |8) ) }

We will now prove the following theorem about the graded polynomial ring $ \cR ( 8|8 ) $. 

\vskip.2cm 

{\bf Theorem 1: }  ${\cal{R}}(8|8)$ is a representation  of $U(\mg)$.

{\vskip 0.25cm}

The elements of $\cS^+_c$ (basis elements for $\mg_c^+$) define the coordinates of the ring $\cR ( 8 |8) $. 
They act on $\mR ( 8 |8) $ by multiplication. The generators $ \mg^+/\mg^+_c$ can be expressed in terms of differential operators which obey the correct commutation relations, proving we have a representation of $U(\mg^+)$. Similarly, we  the generators $ \mg/\mg^+_c$ can be expressed in terms of differential operators which obey the correct commutation relations, proving we have a representation of $U(\mg)$. A straightforward but complicated calculation checks that a correct realization of the psu$(2,2|4)$ generators is provided by (all indices run over 1,2 and repeated indices are summed)
\bea
K^{\dot{\beta}\alpha}&=&-4P_{\alpha_1\dot{\beta}_1}{\partial\over\partial P_{\alpha\dot{\beta}_1}}
{\partial\over\partial P_{\alpha_1\dot{\beta}}}
-4Q^{i+2}{}_\rho {\partial\over\partial Q^{i+2}{}_\alpha}{\partial\over\partial P_{\rho\dot{\beta}}}
-4\bar{Q}_{i\dot{\tau}}{\partial\over\partial\bar{Q}_{i\dot{\beta}}}{\partial\over \partial P_{\alpha\dot{\tau}}}\cr
&&\qquad -4\Delta {\partial\over\partial P_{\alpha\dot{\beta}}}
+8R^{i+2}{}_j\frac{\partial}{\partial\bar{Q}_{j\dot{\beta}}}{\partial\over\partial Q^{i+2}{}_\alpha}
\cr\cr
Q^i{}_\alpha &=& 2 P_{\alpha\dot{\beta}} {\partial\over\partial\bar{Q}_{i\dot{\beta}}}
+ Q^{k+2}{}_\alpha{\partial\over\partial R^{k+2}{}_i}
\cr\cr
\bar{Q}_{i+2\,\,\dot{\alpha}} &=& 2 P_{\alpha\dot{\alpha}} {\partial\over\partial Q^{i+2}{}_\alpha}
-\bar{Q}_{l\dot{\alpha}}{\partial\over\partial R^{i+2}{}_l}
\cr\cr
D&=&P_{\alpha\dot{\alpha}}{\partial\over\partial P_{\alpha\dot{\alpha}}}
+{1\over 2} Q^{k+2}{}_\beta{\partial\over\partial Q^{k+2}{}_\beta}
+{1\over 2} \bar{Q}_{k\dot{\beta}}{\partial\over\partial\bar{Q}_{k\dot{\beta}}}
+\Delta
\cr\cr
M_\alpha{}^\beta&=&P_{\alpha\dot\gamma}{\partial\over\partial P_{\beta\dot\gamma}}
-{1\over 2}\delta^\beta_\alpha P_{\gamma\dot\gamma}{\partial\over\partial P_{\gamma\dot\gamma}}
+Q^{i+2}{}_\alpha{\partial\over\partial Q^{i+2}{}_\beta}
-{1\over 2}\delta^\beta_\alpha Q^{i+2}{}_\gamma {\partial\over\partial Q^{i+2}{}_\gamma} 
\cr\cr
\bar{M}^{\dot\alpha}{}_{\dot{\beta}}&=&
-P_{\gamma\dot\beta}{\partial\over\partial P_{\gamma\dot\alpha}}
+{1\over 2}\delta_{\dot\beta}^{\dot\alpha} P_{\gamma\dot\gamma}{\partial\over\partial P_{\gamma\dot\gamma}}
-\bar{Q}_{i\dot\beta}{\partial\over\partial\bar{Q}_{i\dot{\alpha}}}
+{1\over 2}\delta^{\dot\alpha}_{\dot\beta}\bar{Q}_{i\dot\gamma}{\partial\over\partial\bar{Q}_{i\dot{\gamma}}}
\cr\cr
\bar{S}^{i\dot\alpha}&=&-4P_{\sigma\dot\beta}{\partial\over\partial P_{\sigma\dot\alpha}}
{\partial\over\partial\bar{Q}_{i\dot\beta}}
-2 Q^{l+2}{}_\sigma {\partial\over\partial P_{\sigma\dot\alpha}}{\partial\over\partial R^{l+2}{}_i}
-4 \bar{Q}_{l\dot\beta}{\partial\over\partial\bar{Q}_{l\dot\alpha}}{\partial\over\partial\bar{Q}_{i\dot\beta}}\cr
&&\qquad -4\Delta{\partial\over\partial \bar{Q}_{i\dot\alpha}}
+4R^{j+2}{}_l{\partial\over\partial \bar{Q}_{l\dot\alpha}}{\partial\over\partial R^{j+2}{}_i}\cr\cr
\bar{S}^{i+2\dot{\beta}}&=&-2Q^{i+2}{}_\rho{\partial\over\partial P_{\rho\dot\beta}}
+4 R^{i+2}{}_j{\partial\over\partial\bar{Q}_{j\dot\beta}}
\cr\cr
S_i{}^\alpha &=& 2\bar{Q}_{i\dot\tau}{\partial\over\partial P_{\alpha\dot\tau}}
+4 R^{j+2}{}_i{\partial\over\partial Q^{j+2}{}_\alpha}
\cr\cr
S_{i+2}{}^\alpha&=&4P_{\sigma\dot\beta}{\partial\over\partial P_{\alpha\dot\beta}}{\partial\over\partial Q^{i+2}{}_\sigma}
+4Q^{k+2}{}_\rho{\partial\over\partial Q^{k+2}{}_\alpha}{\partial\over\partial Q^{i+2}{}_\rho}
+4\Delta {\partial\over\partial Q^{i+2}{}_\alpha}\cr
&&\qquad -2\bar{Q}_{j\dot\tau}{\partial\over\partial P_{\alpha\dot\tau}}{\partial\over\partial R^{i+2}{}_j}
-4R^{j+2}{}_k{\partial\over\partial Q^{j+2}{}_\alpha}{\partial\over\partial R^{i+2}{}_k}
\cr\cr
R^{i+2}{}_{j+2}&=&-Q^{i+2}{}_\alpha {\partial\over\partial Q^{j+2}{}_\alpha}
+{\delta^i_j\over 4}Q^{k+2}{}_\alpha {\partial\over\partial Q^{k+2}{}_\alpha}
-{\delta^i_j\over 4}\bar{Q}_{k\dot\gamma}{\partial\over\partial \bar{Q}_{k\dot\gamma}}
-R^{i+2}{}_k{\partial\over\partial R^{j+2}{}_k}+{\Delta\over 2}\delta^i_j
\cr\cr
R^i{}_j&=& \bar{Q}_{j\dot\beta} {\partial\over\partial\bar{Q}_{i\dot\beta}}
-{\delta^i_j\over 4}\bar{Q}_{k\dot\gamma}{\partial\over\partial \bar{Q}_{k\dot\gamma}}
+{\delta^i_j\over 4}Q^{k+2}{}_\alpha {\partial\over\partial Q^{k+2}{}_\alpha}
+R^{k+2}{}_j{\partial\over\partial R^{k+2}{}_i}-{\Delta\over 2}\delta^i_j
\cr\cr
R^i{}_{j+2}&=&2P_{\alpha\dot\rho}{\partial\over\partial Q^{j+2}{}_\alpha}{\partial\over\partial\bar{Q}_{i\dot\rho}}
-Q^{l+2}{}_\alpha{\partial\over\partial R^{l+2}{}_i}{\partial\over\partial Q^{j+2}{}_\alpha}
-\bar{Q}_{l\dot\rho}{\partial\over\partial \bar{Q}_{i\dot\rho}}{\partial\over\partial R^{j+2}{}_l}\cr
&&\qquad +\Delta{\partial\over\partial R^{j+2}{}_i}
-R^{k+2}{}_l{\partial\over\partial R^{k+2}{}_i}{\partial\over\partial R^{j+2}{}_l}
\label{psu224generators}
\eea
This gives a 1/2 BPS representation of $\psu(2,2|4)$. The check that these generators do indeed close the correct algebra (as given in Appendix \ref{psu224})   can be carried out with the help of mathematica. The notebook which performs this check has been included among the source files of this arXiv submission.

Notice that our representation (\ref{psu224generators}) includes the arbitrary parameter $\Delta$. The half-BPS condition is respected because $\Delta$ appears both in the dilatation generator $D$ as well as in the R-symmetry generators and the BPS condition is stated in terms of $D$ and the R-symmetry generators. By setting $\Delta=n$ we obtain the superconformal  $ (\psu(2,2|4) )$ representation constructed on the superconformal primary $Z^n$ for the U$(1)$ theory, which is isomorphic to the superconformal representation constructed on $\Tr (Z^n)$ in the U$(N)$ theory.

\subsection{Super-ring and field theory states}

We will now specialise our considerations to the case where the lowest weight state has dimension $\Delta=1$. The representation  corresponds to the states generated by acting with the super-algebra on the field  $Z$ in the $U(1)$ $\cN=4$ SYM theory. We describe  concretely how  the states of low-dimension  in $V_{\cR}^{(0)}  $ map  to the 
states in $V_{\rm{phys}}^{\rm BPS}$ corresponding (via the operator-state correspondence of CFT) to the 
elementary fields of $ \cN =4$ SYM. The map has a non-trivial kernel, and we give examples of elements in this kernel.  We motivate, using the examples, the need for a resolution in terms of modules in describing $V_{\rm{phys}}^{\rm BPS}$ cleanly. While $V_{\rm{phys}}^{\rm BPS}$ and $V^{(0)}_{\cR}$ are representations of $U(\mg)$ and the map $\varphi_{0*}:V^{(0)}_{\cR}\rightarrow V_{\rm{phys}}^{\rm BPS}$  is equivariant,  typically the $V^{(k)}_{\cR}$ for $k >0$ are not $U(\mg)$ modules and the higher maps are not $U(\mg)$ equivariant. These sequences of modules break the $U(\mg)$ symmetry to a smaller algebra.

The elementary fields of $\cN=4$ super Yang-Mills are six scalars in the ${\tiny\yng(1,1)}$ representation of su(4), a Weyl fermion in the fundamental and one in the anti-fundamental of su(4) as well the field strength $F_{\mu\nu}$. The six scalar fields are recovered by acting on the lowest weight state with the su(4) generators. By acting with the supercharges $Q^{i+2}{}_\alpha$ and $\bar{Q}_{i\dot\alpha}$ $i=1,2$ we generate the fermions and field strengths, completely recovering the fundamental field multiplet. In this way, homogeneous polynomials in our ring correspond to operators in the Yang-Mills theory.

To give the details of the map between homogeneous polynomials in the ring and fields in the CFT, we will follow the conventions of \cite{Ryzhov:2003kk}. The scalar fields are described by an antisymmetric matrix $M_{ij}$ $i,j=1,2,3,4$ which obeys the reality condition
\bea
\bar{M}^{jk}=(M_{jk})^\dagger={1\over 2}\epsilon^{jklm}M_{lm}
\eea
The correspondence between elements of the ring and scalars is
\bea
|Z\rangle_{\rm phys}&=& |M_{34}\rangle_{\rm phys}\qquad
R^3{}_i|Z\rangle_{\rm phys}\,\,\,=\,\,\, |M_{i4}\rangle_{\rm phys}\cr\cr 
R^4{}_i|Z\rangle_{\rm phys}&=& |M_{3i}\rangle_{\rm phys}\qquad
R^3{}_1R^4{}_2|Z\rangle_{\rm phys}\,\,\,=\,\,\, |M_{12}\rangle_{\rm phys}\label{physstates}
\eea
There is some freedom in the correspondence between physical states of the Yang-Mills theory and elements of the ring, arising as a consequence of redundancies where different sequences of transformations of the physical state produce the same result. For example, we also have
\bea
|M_{12}\rangle_{\rm phys}&=&-R^3{}_2R^4{}_1|Z\rangle_{\rm phys}
\eea
which implies the following equality
\bea
R^3{}_1R^4{}_2|Z\rangle_{\rm phys}=-R^3{}_2R^4{}_1|Z\rangle_{\rm phys}\label{scalconst}
\eea
This equality is respected in $V_{ \rm{phys} }^{\rm BPS}$. However, in ${\cal V}^{(0)}_{\cR}$, $R^3{}_1R^4{}_2|Z\rangle_{\cR}$ and $-R^3{}_2R^4{}_1|Z\rangle_{\cR}$ are distinct states. The constraint \eqref{scalconst} is built into the map
\bea
\varphi_{0*}:{\cal V}_{\cal R}^{(0)}\to V^{\rm BPS}_{\rm phys}
\eea
which takes
\bea
\varphi_{0*}(|Z\rangle_{\cal R})= |Z\rangle_{\rm phys}
\eea
as the statement that
\bea
R^3{}_1R^4{}_2|Z\rangle_{\cal R}+R^3{}_2R^4{}_1|Z\rangle_{\cal R}
\eea
is in the kernel of $\varphi_{0*}(\cdot)$. Now consider the fermions. Eight of the fermions are given by 
\bea
Q^3{}_\alpha|Z\rangle_{\rm phys}&=&|\lambda_{4\alpha}\rangle_{\rm phys}\qquad 
Q^4{}_\alpha|Z\rangle_{\rm phys}\,\,\,=\,\,\, -|\lambda_{3\alpha}\rangle_{\rm phys}\cr\cr
\bar{Q}_{1\dot\alpha}|Z\rangle_{\rm phys}&=& |\bar{\lambda}^2_{\dot\alpha}\rangle_{\rm phys}\qquad
\bar{Q}_{2\dot\alpha}|Z\rangle_{\rm phys}\,\,\,=\,\,\,-|\bar{\lambda}^1_{\dot\alpha}\rangle_{\rm phys}
\eea
At first sight these identification may look a little odd since su(4) indices do not  seem to match.
However, recalling that $|Z\rangle_{\rm phys}=|M_{34}\rangle_{\rm phys}$, the first relation says $Q^3{}_\alpha\cdot M_{34}=\lambda_{4\alpha}$, which is su(4) covariant.
Note also that the su(4) representation of $M_{ij}$ is self dual so that $M_{34}={1\over 2}\epsilon_{34jk}M^{jk}=M^{12}$. Consequently $\bar{Q}_{1\dot\alpha}\cdot M_{34}=\bar{Q}_{1\dot\alpha}\cdot M^{12}=\bar{\lambda}^2_{\dot\alpha}$, which is again su(4) covariant. The remaining eight fermions are 
\bea
R^4{}_1 Q^3{}_\alpha|Z\rangle_{\rm phys}&=&|\lambda_{1\alpha}\rangle_{\rm phys}\qquad
R^4{}_2 Q^3{}_\alpha|Z\rangle_{\rm phys}\,\,\,=\,\,\,|\lambda_{2\alpha}\rangle_{\rm phys}\cr\cr 
R^3{}_2\bar{Q}_{1\dot\alpha}|Z\rangle_{\rm phys}&=&|\bar{\lambda}^3_{\dot\alpha}\rangle_{\rm phys}\qquad 
R^4{}_2\bar{Q}_{1\dot\alpha}|Z\rangle_{\rm phys}\,\,\,=\,\,\,|\bar{\lambda}^4_{\dot\alpha}\rangle_{\rm phys}
\eea
There is again some freedom in the correspondence between fermion fields and the ring, arising since, for example
\bea
R^4{}_1 Q^3{}_\alpha|Z\rangle_{\rm phys}=-R^3{}_1 Q^4{}_\alpha|Z\rangle_{\rm phys}\label{ferconst}
\eea
This constraint implies that
\bea
R^4{}_1 Q^3{}_\alpha|Z\rangle_{\cR}+R^3{}_1 Q^4{}_\alpha|Z\rangle_{\cR}
\eea
is in the kernel of $\varphi_{0*}(\cdot)$. The field strength $F_{\mu\nu}$ transforms in the reducible $(1,0)\oplus (0,1)$ irrep of SO(1,3) corresponding to two irreducible components. These are the dual and self dual parts, $F_{\alpha\beta}=F_{\beta\alpha}$ and $F_{\dot\alpha\dot\beta}=F_{\dot\beta\dot\alpha}$, of the field strength in bispinor notation. These are su(4)  scalars. The corresponding physical states are
\bea
\epsilon_{ij}Q^{i+2}{}_\alpha Q^{j+2}{}_\beta|Z\rangle_{\rm phys}\,\,\,
=\,\,\,|F_{\alpha\beta}\rangle_{\rm phys}\qquad
\epsilon^{ij}\bar{Q}_{i\dot\alpha} \bar{Q}_{j\dot\beta}|Z\rangle_{\rm phys}\,\,\,
=\,\,\,|F_{\dot\alpha\dot\beta}\rangle_{\rm phys}
\eea
$\epsilon_{ij}$ is antisymmetric so that, accounting for the Grassmann nature of the supercharges, we see that $\epsilon_{ij}Q^{i+2}{}_\alpha Q^{j+2}{}_\beta$ is symmetric under $\alpha\leftrightarrow\beta$ and 
$\epsilon^{ij}\bar{Q}_{i\dot\alpha} \bar{Q}_{j\dot\beta}$ is symmetric under $\dot\alpha\leftrightarrow\dot\beta$.
This confirms that they are in the correct Lorentz representations.
Using the differential operators introduced in (\ref{psu224generators}) above, with $\Delta=1$, we can verify that 
(no sum on $k$ and $k=1,2,3,4$; $i,j$ are summed over 1,2)
\bea
  R^k{}_k \cdot \epsilon_{ij}Q^{i+2}{}_\alpha Q^{j+2}{}_\beta|Z\rangle_{\rm phys}
=R^k{}_k \cdot \epsilon^{ij}\bar{Q}_{i\dot\alpha} \bar{Q}_{j\dot\beta}|Z\rangle_{\rm phys}=0
\eea
so these are SU(4) scalars, as they must be.

In summary, we have introduced a super polynomial ring $U ( \mg_c^+) = {\cal R}(8|8)$ and have given a realization of psu(2,2$|$4) in
terms of differential operators on the super-ring. Acting on the half-BPS state with  dimension $\Delta=1$, spin (0,0) and su(4) charge (0,1,0), with the super-ring
produces a module  $V_{\cR}^{(0)}$ of the super-ring. There is a map from $V_{\cR}^{(0)}$
to the state space of the half BPS fundamental field representation. The map is not 1-1 because of equations of motion and because of redundancies in expressing the same state 
in the half-BPS representation in multiple ways as a product of elements in $U ( \mg_c^+)$ acting on the half-BPS state. Examples of this second type of redundancy are provided by (\ref{scalconst}) and (\ref{ferconst}). In physical language, we may say that the physical fundamental field state space arises by imposing constraints on  the module  $V_{\cR}^{(0)}$.  A systematic description of the constraints, and further, constraints between constraints and so on, uses the notion of a resolution. A resolution gives the state space of interest, here the half-BPS fundamental field representation,  as an exact sequence of modules of $U ( \mg_c^+) = {\cal R}(8|8)$. In the next section we
will give a complete description for generators of the module of constraints  and show that this is consistent with the known character formula for the half-BPS representation with lowest weight having dimension $\Delta =1$. 


\section{Polynomial (super) rings and characters}\label{polyringsandchars}

In the last section we defined a super-polynomial ring $\mathcal{R}(8|8)$, with eight bosonic and eight fermionic generators and we constructed a realisation of psu$(2,2|4)$ in terms of differential operators. We also defined the module ${\cal V}_{\cal R}^{(0)}$ obtained by allowing the ring to act on the lowest weight state $|Z\rangle_{\cR}$. The map from ${\cal V}_{\cal R}^{(0)}$ to the half-BPS fundamental field representation has a non-trivial kernel. We have described this kernel using the language of constraints to be imposed. In this section we will extend this discussion by giving a complete identification of the kernel using the language of constraints. This list of constraints will be confirmed by computing the associated Hilbert series and the demonstrating that this Hilbert series agrees with the known character formula for the half-BPS fundamental field representation. We start with a discussion of a BGG type resolution for superalgebras which connects to the character formulae in \cite{BDHO}, before moving on to discussion of our super-polynomial ring ${\cal V}_{\cal R}^{(0)}=\mathcal{R}(8|8)$.

\subsection{From super-ring resolutions to character formulae}

The complex scalar $Z$ defines, in radial quantization, a state $|Z\rangle_{\rm{physical}}=|Z\rangle_{\rm{phys}}$. In order to have a clear picture of how the irreducible half-BPS representation arises from the action of the generators of psu$(2,2|4)$ on  this lowest weight state, it is convenient to define a state $|Z\rangle^{(0)}_{\rm{HW}}$ which is used to generate a Verma module $U(\mg^+)|Z\rangle_{\rm{HW}}^{(0)}$. This Verma module has a basis in 1-1 correspondence with the elements of $U(\mg^+)$. Considering the finite dimensional subspaces of $ U(\mg^+)|Z\rangle^{(0)}_{\rm{HW}}$ for fixed eigenvalues of  $\mg_0$, these are isomorphic vector spaces to the eigenspaces of the $\mg_0$ action on $ U(\mg^+)$ by commutators. In other words, there are no relations in $U(\mg^+)|Z \rangle^{(1)}_{\rm{HW}}$ beyond those that are already in $U (\mg^+)$, where we think of $U(\mg^+)$ as a free associative algebra generated by the generators in (\ref{psu224generators}), subject to the commutation relations for this subspace of psu$(2,2|4)$. The physical state space $V_{\rm{phys}}^{BPS} $ is obtained from $ U(\mg^+)|Z\rangle^{(0)}_{\rm{ HW}}$ by setting to zero certain states; these constraints are related to equations of motion. The states being set to zero can themselves be usefully described in terms of a Verma module, subject to other states being set to zero. This whole structure of constraints is encoded in a long exact sequence
\bea\label{BGGsup}   
0\rightarrow {\cal V}_{\rm{HW}}^{(F)}\rightarrow\cdots\rightarrow {\cal V}_{\rm{HW}}^{(1)}\rightarrow {\cal V}_{\rm{HW}}^{(0)}\rightarrow V^{\rm{BPS}}_{\rm{phys}}\rightarrow 0 
\eea
This is a generalization to superalgebras of the BGG resolution for ordinary Lie algebras \cite{superBGG1,superBGG2,superBGG3}. It is expected to underly the character formulae in \cite{BDHO}. We will refer to the lowest weight state in the Verma module $k$ steps away from the physical state space as $|Z^{(k-1)}\rangle_{\rm{HW}}$.

In the case of the  $V^{\rm{BPS}}_{\rm{phys}}$, we have the special property that the physical lowest weight state $|Z\rangle_{\rm{phys}}$ is annihilated, by virtue of being a lowest weight state, by all the elements $a_-\in\mg^{-}$. Further, the BPS condition means that for any element ($\mg_+$ is defined in (\ref{ingcplus}))
\bea 
a_+' |Z \rangle_{ \rm{phys} } = 0 ~~~~ \hbox{for all} ~~~~ a_+' \in \mg_+ 
~~~\hbox{and} ~~~ a_+' \notin\mg_c^+
\eea 
The rather special property of the ${\cal N}=4$ half-BPS representation is that all the generators of $\mg^+$ which do not annihilate $|Z\rangle_{\rm{phys}}$ form a graded-commutative Lie sub-algebra. This suggests that the structure of $V^{\rm{BPS}}_{\rm{phys}}$ can be usefully studied, in analogy to the exact sequence in \eqref{BGGsup}, by using free modules over the graded-commutative ring instead of Verma modules of the form $U(\mg^+)|Z^{(k)}\rangle_{\rm{HW}}$. Exact sequences of modules of commutative rings have been very useful in the combinatorics of composite operators in free scalar field theory \cite{PrimsCC}. We will write these modules of  $\cR(8|8)=U(\mg_c^+)$  as  ${\cal V}_{\cR }^{(k)}=U( \mg^+_c)| Z^{(k)}\rangle_{\cR}$, thus producing exact sequences of the form 
\bea\label{moduleres}  
 0 \rightarrow {\cal V}_{\cR }^{ (F) } \rightarrow \cdots 
\rightarrow {\cal V}_{ \cR  }^{(1)}\rightarrow {\cal V}_{ \cR  }^{(0)}  \rightarrow 
V^{ \rm{BPS} }_{ \rm{phys}   } \rightarrow 0 
\eea
The states $|Z^{(0)}\rangle_{\cR}$ have properties analogous to $|Z\rangle_{\rm phys}  $, but with some differences. With
\bea 
a_0 \in \mg_0,\qquad a_+ \in \mg_+,\qquad a_+^c \in \mg_c^+,\qquad  
a_+^\prime\in\mg_+\quad a_+^\prime\notin\mg_+^c 
\eea
we have 
\bea 
&& a_0|Z^{(0)}\rangle_{\cR} = a_0(Z) |Z^{(0)}\rangle_{\cR} \cr 
&& a_- |Z^{(0)}\rangle_{\cR} = 0 \qquad
a_+^\prime|Z^{(0)}\rangle_{\cR}\ne 0 
\eea
and 
\bea 
&& a_0 | Z\rangle_{\rm phys} = a_0(Z) |Z\rangle_{ \rm phys} \cr 
&& a_- | Z\rangle_{\rm phys} =0 \qquad
a_+^\prime |Z\rangle_{\rm phys} =  0 
\eea
Here $a_0$ runs over the elements of the Cartan subalgebra. Using the differential operators (\ref{psu224generators}) it is simple to verify that
\bea
D| Z\rangle_{\rm phys}&=& |Z\rangle_{\rm phys}\qquad
H_1| Z\rangle_{\rm phys}\,\,\,=\,\,\,0\qquad
H_2| Z\rangle_{\rm phys}\,\,\,=\,\,\,| Z\rangle_{\rm phys}\cr\cr
H_3| Z\rangle_{\rm phys}&=&0\qquad
J_3| Z\rangle_{\rm phys}\,\,\,=\,\,\,0\qquad
\bar{J}_3| Z\rangle_{\rm phys}\,\,\,=\,\,\,0
\eea
which implies
\bea
a_0(D)=1\qquad a_0(H_2)=1,\qquad a_0(H_1)=a_0(H_3)=a_0(J_3)=a_0(\bar{J}_3)=0
\eea

We will now list the constraints that must be imposed to pass from the ring $V_{\cR}^{(0)}=  \cR(8|8)$ to the half BPS representation $V^{\rm{BPS}}_{\rm{phys}}$. These constraints are related to equations of motion, to  kinematic relations due to the fact that the same physical state can be obtained by acting with different sequences of superalgebra generators  on the lowest weight state and to relations that follows as a consequence of the psu$(2,2|4)$ algebra. We find that a total of 64 constraints must be imposed on $V_{\cR}^{(0)}$ to recover $V_{\rm{phys}}^{\rm BPS}$. All constraints are quadratic. The constraints are

\begin{align}
R^{k+2}{}_{i}R^{k+2}{}_{j}|\phi_{34}\rangle^{(0)}&=0\qquad k=1,2,\,\,\,(i,j)=(1,1),(1,2),(2,2) &[6]\cr\cr
R^3{}_{i}R^4_{i}|\phi_{34}\rangle^{(0)}&=0\qquad i=1,2&[2]\cr\cr
R^3{}_{1}R^4{}_{2}|\phi_{34}\rangle^{(0)}+R^3{}_{2}R^4{}_{1}|\phi_{34}\rangle^{(0)}&=0&[1]\cr\cr
R^{k+2}{}_iQ^{k+2}{}_\alpha|\phi_{34}\rangle^{(0)}&=0\qquad k=1,2,\,\,\, i=1,2\,\,\,\alpha=1,2&[8]\cr\cr
R^4{}_iQ^3{}_\alpha|\phi_{34}\rangle^{(0)}+R^3{}_iQ^4{}_\alpha|\phi_{34}\rangle^{(0)}&=0\qquad i=1,2\,\,\,\alpha=1,2&[4]\cr\cr
R^{k+2}{}_i\bar{Q}_{i\dot\alpha}|\phi_{34}\rangle^{(0)}&=0\qquad i=1,2\,\,\,k=1,2\,\,\,\dot{\alpha}=1,2
&[8]\cr\cr
R^{k+2}{}_1\bar{Q}_{2\dot\alpha}|\phi_{34}\rangle^{(0)}+R^{k+2}{}_2\bar{Q}_{1\dot{\alpha}}|\phi_{34}\rangle^{(0)}&=0\qquad k=1,2\,\,\,\dot{\alpha}=1,2&[4]\cr\cr
\bar{Q}_{11}\bar{Q}_{22}|\phi_{34}\rangle^{(0)}+\bar{Q}_{21}\bar{Q}_{12}|\phi_{34}\rangle^{(0)}&=0&[1]\cr\cr
Q^3{}_1 Q^4{}_2|\phi_{34}\rangle^{(0)}+Q^4{}_1 Q^3{}_2|\phi_{34}\rangle^{(0)}&=0&[1]\cr\cr
Q^{k+2}{}_\alpha\bar{Q}_{j\dot{\alpha}}|\phi_{34}\rangle^{(0)}&=0\qquad j,k,\alpha,\dot{\alpha}=1,2
&[16]\cr\cr
\bar{Q}_{i1}\bar{Q}_{i2}|\phi_{34}\rangle^{(0)}&=0\qquad i=1,2&[2]\cr\cr
Q^{k+2}{}_1Q^{k+2}{}_2|\phi_{34}\rangle^{(0)}&=0\qquad k=1,2&[2]\cr\cr
P^{\alpha\dot\alpha}P_{\alpha\dot\alpha}|\phi_{34}\rangle^{(0)}&=0 &[1]\cr\cr
\epsilon^{\alpha\beta}P_{\alpha\dot{\alpha}}Q^{k+2}_\beta|\phi_{34}\rangle^{(0)}&=0\qquad k,\dot\alpha=1,2 &[4]\cr\cr
\epsilon^{\dot\alpha\dot\beta}P_{\alpha\dot{\alpha}}\bar{Q}_{k\dot\beta}|\phi_{34}\rangle^{(0)}&=0\qquad k,\alpha=1,2 &[4]
\label{HBS}
\end{align}
We have listed the number of constraints in square brackets on the right. The first, second, fourth, sixth, eleventh and twelfth lines are statements true in the antisymmetric ${\tiny\yng(1,1)}$ su(4) representation to which the 6 scalars belong. The third, fifth, seventh, eighth and ninth lines identify the same state, obtained by acting with different sequences of superalgebra generators. The tenth line is a property of the anticommutator between $Q^{k+2}{}_\alpha$ and $\bar{Q}_{j\dot{\alpha}}$, as well as a property of the half BPS representation. The last three lines correspond to the massless Klein-Gordon and massless Dirac equations. This list of constraints looks incomplete, since Maxwell's equations do not appear in the above list. Maxwell's equations ($F_{\mu\nu}$ is the field strength tensor and $*F_{\mu\nu}$ is the Hodge dual of the field strength tensor)
\bea
\partial^\mu F_{\mu\nu}=0\qquad \partial^\mu *F_{\mu\nu}=0
\eea
which, in bispinor notation, are
\bea
\epsilon^{\gamma\alpha}P_{\gamma\dot{\gamma}}F_{\alpha\beta}
&=&\epsilon^{\gamma\alpha}P_{\gamma\dot{\gamma}}Q^3{}_\alpha Q^4{}_\beta
-\epsilon^{\gamma\alpha}P_{\gamma\dot{\gamma}}Q^4{}_\alpha Q^3{}_\beta\cr\cr
\epsilon^{\gamma\alpha}P_{\gamma\dot{\gamma}}\tilde{F}_{\dot\alpha\dot\beta}&=&
\epsilon^{\gamma\alpha}P_{\gamma\dot{\gamma}}\bar{Q}_{1\dot\alpha} \bar{Q}_{2\dot\beta}
-\epsilon^{\gamma\alpha}P_{\gamma\dot{\gamma}}Q_{2\dot\alpha}\bar{Q}_{1\dot\beta}
\eea
are implied by the Dirac equation, which appear in the constraints. To write the equations above, we have used the identification of the dual and anti-self dual components of the field strength tensor, developed in the previous section.

With these constraints in hand, it is now possible to state a key result

{\vskip 0.25cm}

\noindent
{\bf Theorem:} The resolution (\ref{moduleres}) with ${\cal V}^{(1)}$ defined by the constraints given in (\ref{HBS}) has a coarsened character that agrees with the coarsened character of the half BPS irrep computed in \cite{BDHO}.

{\vskip 0.25cm}

\noindent
This theorem is a convincing test of the constraints we have identified. To prove the theorem, we begin by computing the Hilbert series using the following sage code
\begin{verbatim}
names = '(r13,r14,r23,r24,puu,pud,pdu,pdd,q3u,q3d,q4u,q4d,q1bu,q1bd,q2bu,q2bd)'
Igens = ('r13*r13','r13*r23','r23*r23','r14*r14','r14*r24','r24*r24','r13*r14',
              'r23*r24','r13*r24+r23*r14','r13*q3u','r23*q3u','r14*q4u','r24*q4u',
              'r14*q3u+r13*q4u','r24*q3u+r23*q4u','r13*q3d','r23*q3d','r14*q4d',
              'r24*q4d','r14*q3d+r13*q4d','r24*q3d+r23*q4d','r13*q1bu','r23*q2bu',
              'r14*q1bu','r24*q2bu','r14*q2bu+r24*q1bu','r13*q2bu+r23*q1bu',
              'r13*q1bd','r23*q2bd','r14*q1bd','r24*q2bd','r14*q2bd+r24*q1bd',
              'r13*q2bd+r23*q1bd','q1bu*q2bd+q2bu*q1bd','q3u*q4d+q4u*q3d',
              'q1bu*q3u','q1bu*q3d','q1bd*q3u','q1bd*q3d','q2bu*q3u','q2bu*q3d',
              'q2bd*q3u','q2bd*q3d','q1bu*q4u','q1bu*q4d','q1bd*q4u','q1bd*q4d',
              'q2bu*q4u','q2bu*q4d','q2bd*q4u','q2bd*q4d','q2bu*q2bd','q1bu*q1bd',
              'q3u*q3d','q4u*q4d','puu*pdd-pud*pdu','puu*q3d-pdu*q3u',
              'pud*q3d-pdd*q3u','puu*q4d-pdu*q4u','pud*q4d-pdd*q4u',
              'pud*q1bu-puu*q1bd','pdd*q1bu-pdu*q1bd','pud*q2bu-puu*q2bd',
              'pdd*q2bu-pdu*q2bd',
R = singular.ring(0,names,'dp')
A = singular.superCommutative(9,16)
A.set_ring()
I = singular.ideal(Igens)
from sage.rings.polynomial.hilbert import hilbert_poincare_series
hilbert_poincare_series(I)
\end{verbatim}
The result is
\bea
{t^3 - 19t^2 - 13t - 1\over (t-1)^3}\label{sageoutput}
\eea
This is the coarsened character of the resolution. As a test of this result, we can verify that the RHS of (\ref{sageoutput}) has positive expansion coefficients when expanded as a power series in $t$. This must be the case since these coefficients are counting states in the 1/2 BPS multiplet produced by acting with a certain number of generators. To understand the positivity of the RHS we study the LHS
\bea
{\rm LHS}=A_1+A_2+A_3
\eea
with
\bea
A_1&=&{(1+4t+t^2)(1-t^2)\over (1-t)^4}
={(1+4t+t^2)(1+t)\over (1-t)^3}\cr\cr
&=&(1+4t+t^2)(1+t)(\sum_{n=0}^\infty t^n)^3
\eea
which has manifestly positive expansion coefficients, and
\bea
A_2&=&{2\cdot 4t(1+t)(1-t)\over (1-t)^4}
={8t(1+t)\over (1-t)^3}\cr\cr
&=&8t(1+t)(\sum_{n=0}^\infty t^n)^3
\eea
which is again manifestly positive, and
\bea
A_3&=&{2(3t^2-4t^3+t^4)\over (1-t)^4}
={2t^2(t-1)(t-3)\over (1-t)^4}\cr\cr
&=&{2t^2(1-t+2)\over (1-t)^3}
={4t^2\over (1-t)^3}+{2t^2\over (1-t)^2}\cr\cr
&=&4t^2(\sum_{n=0}^\infty t^n)^3+2t^2(\sum_{n=0}^\infty t^n)^2
\eea
which is again manifestly positive. To complete the proof of the theorem we need to show that it reproduces the character for the half BPS representation, after that character is rewritten using the grading adopted here. This rewriting gives the coarsened character of the half BPS irrep. Recall that the character is \cite{BDHO}
\begin{eqnarray}
\chi^{{1\over 2},{1\over 2}}_{(1;0,1,0;0,0)}(s,x,\bar{x},u)&=&{\cal D}_0 (s,x,\bar x)\chi_{(1,1,0,0)}(u)\cr
&+&{\cal D}_{1\over 2} (s,x,\bar x)\chi_{(1,0,0,0)}(u)+\bar{\cal D}_{1\over 2} (s,x,\bar x)\chi_{(1,1,1,0)}(u)\cr
&+&{\cal D}_{1} (s,x,\bar x)+\bar{\cal D}_{1} (s,x,\bar x)
\end{eqnarray}
The su(4) characters are
\begin{eqnarray}
\chi_{(1,0,0,0)}&=&u_1+u_2+u_3+u_4\qquad
\chi_{(1,1,1,0)}\,\,\,=\,\,\,u_1^{-1}+u_2^{-1}+u_3^{-1}+u_4^{-1}\cr
\chi_{(1,1,0,0)}&=&u_1u_2+u_1u_3+u_1u_4+u_2u_3+u_2u_4+u_3u_4
\end{eqnarray}
The conformal characters are
\begin{eqnarray}
{\cal D}_0(s,x,\bar x) &=& s^2(1-s^4)P(s,x,\bar x)\cr
{\cal D}_{1\over 2}(s,x,\bar x) &=& s^3(\chi_{1\over 2}(x)-s^2\chi_{1\over 2}(\bar x))P(s,x,\bar x)\cr
\bar{\cal D}_{1\over 2}(s,x,\bar x) &=& s^3(\chi_{1\over 2}(\bar x)-s^2\chi_{1\over 2}(x))P(s,x,\bar x)\cr
{\cal D}_1(s,x,\bar x) &=& s^4(\chi_{1}(x)-s^2\chi_{1\over 2}(x)\chi_{1\over 2}(\bar x)+s^4)P(s,x,\bar x)\cr
\bar{\cal D}_1(s,x,\bar x) &=& s^4(\chi_1(\bar x)-s^2\chi_{1\over 2}(x)\chi_{1\over 2}(\bar x)+s^4)P(s,x,\bar x)
\label{DC}
\end{eqnarray}
where
$$
P(s,x,\bar x)={1\over (1-s^2x\bar x)(1-s^2x^{-1}\bar x)(1-s^2x\bar x^{-1})(1-s^2x^{-1}\bar x^{-1})}
$$
The su(2) characters are
\bea
\chi_{1\over 2}=x+{1\over x}\qquad \chi_1(x)=x^2+1+{1\over x^2}
\eea
with $x$ and $\bar{x}$ for the two su(2)s in su(2)$\otimes$su(2). Writing this using the grading adopted in our Hilbert series computation, we have
\begin{eqnarray}
{\cal D}_0\chi_{(1,1,0,0)} &=& (1+4t+t^2)(1-t^2)P(t)\cr
{\cal D}_{1\over 2}\chi_{(1,0,0,0)} &=& (4t+4t^2)(1-t)P(t)\cr
\bar{\cal D}_{1\over 2}\chi_{(1,1,1,0)} &=& (4t+4t^2)(1-t)P(t)\cr
{\cal D}_1 &=& (3t^2-4t^3+t^4)P(t)\cr
\bar{\cal D}_1 &=& (3t^2-4t^3+t^4)P(t)
\end{eqnarray}
where
$$
P(t)={1\over (1-t)^4}
$$
A simple computation now shows that
\bea
&&{(1+4t+t^2)(1-t^2)+2(4t+4t^2)(1-t)+2(3t^2-4t^3+t^4)\over (1-t)^4}\cr\cr
&&={t^3 - 19t^2 - 13t - 1\over (t-1)^3}\label{HalfPF}
\eea
In complete agreement with (\ref{sageoutput}), which completes the proof of the theorem. This gives a detailed and intricate test of the 64 constraints that must be imposed on $V_{\cR}^{(0)}$ to recover $V_{\rm{phys}}^{\rm BPS}$.

At this stage the computation of the resolution appears to be too complex to be handled by Sage. However, by executing the code 
\begin{verbatim}
names = '(r13,r14,r23,r24,puu,pud,pdu,pdd,q3u,q3d,q4u,q4d,q1bu,q1bd,q2bu,q2bd)'
Igens = ('r13*r13','r13*r23','r23*r23','r14*r14','r14*r24','r24*r24','r13*r14',
              'r23*r24','r13*r24+r23*r14','r13*q3u','r23*q3u','r14*q4u','r24*q4u',
              'r14*q3u+r13*q4u','r24*q3u+r23*q4u','r13*q3d','r23*q3d','r14*q4d',
              'r24*q4d','r14*q3d+r13*q4d','r24*q3d+r23*q4d','r13*q1bu','r23*q2bu',
              'r14*q1bu','r24*q2bu','r14*q2bu+r24*q1bu','r13*q2bu+r23*q1bu',
              'r13*q1bd','r23*q2bd','r14*q1bd','r24*q2bd','r14*q2bd+r24*q1bd',
              'r13*q2bd+r23*q1bd','q1bu*q2bd+q2bu*q1bd','q3u*q4d+q4u*q3d',
              'q1bu*q3u','q1bu*q3d','q1bd*q3u','q1bd*q3d','q2bu*q3u','q2bu*q3d',
              'q2bd*q3u','q2bd*q3d','q1bu*q4u','q1bu*q4d','q1bd*q4u','q1bd*q4d',
              'q2bu*q4u','q2bu*q4d','q2bd*q4u','q2bd*q4d','q2bu*q2bd','q1bu*q1bd',
              'q3u*q3d','q4u*q4d','puu*pdd-pud*pdu','puu*q3d-pdu*q3u',
              'pud*q3d-pdd*q3u','puu*q4d-pdu*q4u','pud*q4d-pdd*q4u',
              'pud*q1bu-puu*q1bd','pdd*q1bu-pdu*q1bd','pud*q2bu-puu*q2bd',
              'pdd*q2bu-pdu*q2bd',
R = singular.ring(0,names,'dp')
A = singular.superCommutative(9,16)
A.set_ring()
I = singular.ideal(Igens)
singular.res(I,3)
\end{verbatim}
we are able to generate the first three syzygy modules. The first syzygy module is generated from 64 quadratic constraints - matching the constraints that we listed above. The second syzygy module is generated from 489 constraints, 480 of which are cubic and 9 of which are quartic. The third syzygy module is constructed on 2201 constraints, 2057 of which are quartic and 144 are quintic.

\subsection{so(4,2) representations from subrings of super-rings}

While the complete resolution is too large to be instructive, there are smaller toy problems that can be considered. For example, one can consider the sub ring $\mathbb{C}[R^3{}_1,R^3{}_2,R^4{}_1,R^4{}_2]$, keeping only the first three constraints from our list (\ref{HBS}). The result is a resolution of the ${\tiny\yng(1,1)}$ representation of su(4). Another example is to consider the sub super-ring $\mathbb{C}[R^3{}_1,R^3{}_2,R^4{}_1,R^4{}_2,Q^3{}_1,Q^3{}_2,Q^4{}_1,Q^4{}_2]$, keeping the first five, ninth and twelfth constraints, we obtain a resolution for a state space with a boson in the ${\tiny\yng(1,1)}$ representation of SU(4), a Weyl fermion in the ${\tiny\yng(1)}$ of su(4) and a self dual field strength tensor to give a total of 17 states. These examples are easily explored by modifying the codes we have given so they will not be discussed further. See Appendix \ref{sagesubring}.

The conformal characters themselves can be related to so(2,4) equivariant resolutions, with the appearance of syzygies reflecting the equations of motion. A particularly instructive example is provided by the field strength tensor. For the field strength the character is \cite{Dobrev:2004tk,Dolan:2005wy}

\bea
\chi&=&s^4(\chi_{1}(x)-s^2\chi_{1\over 2}(x)\chi_{1\over 2}(\bar{x})+s^4) P(s,x,\bar{x})\cr\cr
&&+s^4(\chi_{1}(\bar{x})-s^2\chi_{1\over 2}(x)\chi_{1\over 2}(\bar{x})+s^4) P(s,x,\bar{x})
\label{FSChar}
\eea
The first line is from the self dual piece and the second from the anti-self dual piece. The self dual and anti-self dual parts of the field strength tensor are complex linear combinations of the field strength $F_{\mu\nu}$ and it's Hodge dual. The relevant module is generated by the field strength primary over the commutative ring $\mathbb{C}[P^\mu]$. The field strength tensor is a primary operator of dimension 2 in the reducible representation $(1,0)\oplus (0,1)$ of the Lorentz group. This dimension saturates the unitarity bound so that there will be null descendants. These are nothing but the equations of motion, which are the vanishing of a level 1 descendent
\bea
P^\nu F_{\nu\mu}=0
\eea
There is a relation between the constraints, which reads
\bea
P^\mu P^\nu F_{\nu\mu}=0
\eea
In contrast to the equations of motion, this is a relation between constraints and it holds simply as a consequence of the fact that $F_{\nu\mu}$ is antisymmetric. There are identical equations for the Hodge dual ${}^*\!F_{\nu\mu}$. The exact sequence relevant for the field strength representation is given by
\bea
\begin{tikzcd}
  0 \arrow[r," "] & {\cal V}^{(2)}  \arrow[r, "\varphi_{21}"] & {\cal V}^{(1)} \arrow[r, "\varphi_{10}"] & {\cal V}^{(0)}  \arrow[r,"\varphi_{0*}"] & V^{\rm phys} \arrow[r," "] & 0
\end{tikzcd}
\label{FSresolution}
\eea
The module ${\cal V}^{(0)}$ is a direct sum of six modules, generated by the six states $|F_{\mu\nu}\rangle^{(0)}$. The module ${\cal V}^{(1)}$ is a direct sum of eight modules, generated by the eight states $\{|P^\mu F_{\mu\nu}\rangle^{(1)},|P^\mu\, {}^*\!F_{\mu\nu}\rangle^{(1)}\}$. The module ${\cal V}^{(2)}$ is generated by the pair of states $\{|P^\nu |P^\mu F_{\mu\nu}\rangle^{(1)}\rangle^{(2)},|P^\nu |P^\mu \,{}^*\!F_{\mu\nu}\rangle^{(1)}\rangle^{(2)}\}$. The map $\varphi_{0*}$ acts as
\bea
\varphi_{0*}(P_{\alpha_1}\cdots P_{\alpha_k}|F_{\mu\nu}\rangle^{(0)})&=&
P_{\alpha_1}\cdots P_{\alpha_k}|F_{\mu\nu}\rangle^{\rm phys}
\eea
The map $\varphi_{10}$ acts as
\bea
\varphi_{10}(P_{\alpha_1}\cdots P_{\alpha_k}|P^\mu F_{\mu\nu}\rangle^{(1)} )
&=&P_{\alpha_1}\cdots P_{\alpha_k}P^\mu |F_{\mu\nu}\rangle^{(0)}\cr\cr
\varphi_{10}(P_{\alpha_1}\cdots P_{\alpha_k}|P^\mu\,{}^*\! F_{\mu\nu}\rangle^{(1)} )
&=&P_{\alpha_1}\cdots P_{\alpha_k}P^\mu \epsilon_{\mu\nu\alpha\beta}|F^{\alpha\beta}\rangle^{(0)}
\eea
The state $P^\mu |F_{\mu\nu}\rangle^{(0)}$ is a null state. For this state we have
\bea
\varphi_{10}(|P^\mu F_{\mu\nu}\rangle^{(1)} )
=P^\mu |F_{\mu\nu}\rangle^{(0)}\qquad
\varphi_{0*}(P^\mu |F_{\mu\nu}\rangle^{(0)})=0
\eea
There is a similar statement for  $\epsilon_{\nu\mu\alpha\beta}P^\mu |F^{\alpha\beta}\rangle^{(0)}$. The map $\varphi_{21}$ acts as
\bea
\varphi_{21}(P_{\alpha_1}\cdots P_{\alpha_k}|P^\mu |P^\nu F_{\mu\nu}\rangle^{(1)}\rangle^{(2)})&=& 
P_{\alpha_1}\cdots P_{\alpha_k}P^\nu |P^\mu F_{\mu\nu}\rangle^{(1)}\cr\cr
\varphi_{21}(P_{\alpha_1}\cdots P_{\alpha_k}|P^\mu |P^\nu \,{}^*\!F_{\mu\nu}\rangle^{(1)}\rangle^{(2)})&=& P_{\alpha_1}\cdots P_{\alpha_k}P^\nu |P^\mu\,{}^*\! F_{\mu\nu}\rangle^{(1)}
\eea
Note that
\bea
\varphi_{21}(|P^\mu |P^\nu F_{\mu\nu}\rangle^{(1)}\rangle^{(2)})= 
P^\nu |P^\mu F_{\mu\nu}\rangle^{(1)}\qquad
\varphi_{10}(P^\nu |P^\mu F_{\mu\nu}\rangle^{(1)})
=P^\nu P^\mu |F_{\mu\nu}\rangle^{(0)}=0
\eea
$P^\nu |P^\mu F_{\mu\nu}\rangle^{(1)}$ is a null vector in ${\cal V}^{(1)}$ and it maps to the zero vector in ${\cal V}^{(0)}$. It is obvious that there are analogous statements for  $P^\nu |P^\mu\,{}^*\!F_{\mu\nu}\rangle^{(1)}$.

It is easy to verify that $P^\mu |F_{\mu\nu}\rangle^{(0)}$ is null, or equivalently, that it is primary. In the conventions of \cite{CFT4TFT2} we have
\bea
[K_\mu,P_\nu]&=&2M_{\mu\nu}-2\eta_{\mu\nu}D\cr\cr
[M^{\mu\nu},P_\rho]&=&\delta_\rho^\nu \eta^{\mu\tau}P_\tau
-\delta_\rho^\mu \eta^{\nu\tau}P_\tau
\eea
The second commutator sets our conventions for how a vector index transforms. We now easily find
\bea
K^\alpha P^\mu |F_{\mu\nu}\rangle^{(0)}
&=& 2M^{\alpha\mu}|F_{\mu\nu}\rangle^{(0)}
-2\eta^{\alpha\mu}D|F_{\mu\nu}\rangle^{(0)}\cr\cr
&=&2(\delta^\mu_\mu \eta^{\alpha\rho}|F_{\rho\nu}\rangle^{(0)}
-\delta^\alpha_\mu\eta^{\mu\rho} |F_{\rho\nu}\rangle^{(0)}
+\delta^\mu_\nu \eta^{\alpha\rho}|F_{\mu\rho}\rangle^{(0)}
-\delta^\alpha_\nu\eta^{\mu\rho} |F_{\mu\rho}\rangle^{(0)})\cr\cr
&&-4\eta^{\alpha\mu}|F_{\mu\nu}\rangle^{(0)}\cr\cr
&=&0
\eea
Next, consider the state $|X_\nu\rangle^{(1)}\equiv |P^\mu F_{\mu\nu}\rangle^{(1)}$. We want to verify that the descendant $P^\mu |X_\mu\rangle^{(1)}$ is null i.e. that it is primary. A simple computation shows
\bea
K^\alpha P^\mu |X_\mu\rangle^{(1)}
&=& 2M^{\alpha\mu}|X_\mu\rangle^{(1)}
-2\eta^{\alpha\mu}D|X_\mu\rangle^{(1)}\cr\cr
&=&2(\delta^\mu_\mu\eta^{\alpha\rho}|X_\rho\rangle^{(1)}
-\delta^\alpha_\mu\eta^{\mu\rho}|X_\rho\rangle^{(1)})
-6\eta^{\alpha\mu}|X_\mu\rangle^{(1)}\cr\cr
&=&0
\eea
In exactly the same way we can prove that $\epsilon_{\mu\nu\alpha\beta}P^\nu |F^{\alpha\beta}\rangle^{(0)}$ and $|W_\nu\rangle^{(1)}\equiv |P^\mu\,{}^*\!F_{\mu\nu}\rangle^{(1)}$ are null, i.e. that they are primary. The fact that these are primary implies that $\varphi_{21}$, $\varphi_{10}$ and $\varphi_{0*}$ are so(4,2) equivariant maps. Thus, the resolution in (\ref{FSresolution}) explains the character of the field strength multiplet (\ref{FSChar}) and is so(4,2) equivariant.

\section{Homogeneous  and  emergent symmetries in oscillator constructions }\label{emergentsymm}

The super-polynomial ring $U(\mg_c^+)=\cR(8|8)$ that we have used to give a resolution of the fundamental field representation of psu$(2,2|4)$ naturally defines a Fock space associated to $8$ bosonic and $8$ fermionic oscillators. superalgebra realizations using bosonic and fermionic oscillators have had many applications in super Yang-Mills theory and in AdS/CFT \cite{Beisert:2003jj,Dobrev:2018lsx,wittenoscilattor}. The Fock space at hand exhibits a hidden u$(8|8)$ symmetry which is present in the ${\cal N}=4$ super Yang-Mills. Any state can be written as a polynomial in the creation operators, acting on the Fock vacuum. This is a homogeneous symmetry: it only mixes states constructed from polynomials of a fixed degree. In previous sections we have explained how to obtain the psu$(2,2|4)$ symmetry by imposing a sequence of constraints. In contrast to u$(8|8)$, the action of psu$(2,2|4)$ is not homogenous and polynomials of different degree mix. The ring $\cR(8|8)$ is a representation of the psu$(2,2|4)$ symmetry as well as of the u$(8|8)$ symmetry. The irrep $V$ is a representation of psu$(2,2|4)$ but not of u$(8|8)$. The higher syzygies are not representations of psu$(2,2|4)$, but they are representations of u$(8|8)$. This section gives a physical algorithm for constructing the syzygies, using the oscillator Fock spaces, with the natural inner product (\ref{InnerProduct}). There are of course well-developed algorithms that compute syzygies, based on Groebner bases and they are implemented in Sage and Singular \cite{Cox}. These are probably more efficient than the algorithms that we outline here. The key point of our discussion is to show that the hidden symmetries, such as the homogeneous u$(8|8)$ symmetry, and their associated inner products, are useful in giving a physical formulation of the problem of constructing first order as well as higher order syzygies. By contrast, the non-homogeneous emergent psu$(2,2|4)$ symmetry is associated with a psu$(2,2|4)$ invariant inner product of the conformal field theory. In addition, an implementation of our algorithm in simple cases serves as a simple physical consistency check on the results coming from the mathematical syzygy algorithm. By employing a Fock space description, in terms of which syzygies are represented as states in the Fock space, we translate results from the language of rings and modules into the language of quantum mechanics. This language illustrates how one can obtain quantum mechanical systems in terms of an ideal constructed from oscillators, leading to a new symmetry on the quotient. 

It will be helpful to introduce some notation for the discussion that follows. Our resolution has the form
\bea\label{resHalfBPS} 
0\longrightarrow {\cal V}^{(n)} \longrightarrow \cdots
\longrightarrow {\cal V}^{(2)} \longrightarrow {\cal V}^{(1)}
\longrightarrow {\cal V}^{(0)} =  \cR(8|8) \longrightarrow V\longrightarrow 0
\eea
Here ${\cal V}^{(0)}$ is the ring ${\cal R}(8|8)$ itself, while $V$ is the irrep being resolved. We say module ${\cal V}^{(k)}$ has depth $k$. Recall that the image of the generators of the module ${\cal V}^{(i)}$ obey a set of relations. These relations are in correspondence with the syzygies that generate the module ${\cal V}^{(i+1)}$. For example, the constraints that determine the kernel of the map from the ring $\cR(8|8)$ to the irrep $V$, are associated to the first syzygies, which define the module ${\cal V}^{(1)}$. The sequence terminates because there are no relations obeyed by the image of the generators of the module ${\cal V}^{(n)}$, or equivalently, because the kernel of the map from ${\cal V}^{(n)}$ to ${\cal V}^{(n-1)}$ is trivial. At a given depth, syzygies of different degrees can appear. The syzygy at depth $i$ which has degree $d$ is denoted by
\bea
g^{(i),[d]}_a
\eea
with the index $a$ labelling the different syzygies.  

\subsection{Oscillator Fock space and u$(8|8)$ symmetry}

The super-polynomial ring ${\cal V}_{\cR}^{(0)}$ that we have used to construct a realization of psu$(2,2|4)$ has $8$ bosonic
\bea 
P_1,\,\, P_2,\,\, P_3,\,\, P_4,\,\, R^3_1,\,\, R^3_2,\,\, R^4_1,\,\, R^4_2  
\eea
and $8$ fermionic generators
\bea 
Q^3_ 1,\,\, Q^3_2,\,\, Q^4_1,\,\, Q^4_2,\,\, \bar{Q}^1_1,\,\, \bar{Q}^1_2,\,\, 
\bar{Q}^2_1,\,\, \bar{Q}^2_2 
\eea
In constructing the psu$(2,2|4)$ algebra, we use derivatives with respect to these coordinates of the ring. The fermionic derivatives have the usual properties of fermions: they square to zero, and distinct fermionic derivatives anti-commute with each other, while commuting with the bosonic variables and bosonic derivatives. These are precisely the relations of the superalgebra of 8 bosonic oscillators and 8 fermionic oscillators. We can identify 
\bea 
\{P_1,\,P_2,\,P_3,\,P_4,\,R^3_1,\,R^3_2,\,R^4_1,\,R^4_2\} 
\leftrightarrow \{a_1^\dagger,\,a_2^\dagger,\,a_3^\dagger,\, a_4^\dagger,\, 
a_5^\dagger,\,a_6^\dagger,\,a_7^\dagger,\, a_8^\dagger\} 
\eea
The corresponding derivatives are identified with the annihilation operators. We can also identify
\bea 
\{Q^3_ 1,\,Q^3_2,\,Q^4_1,\,Q^4_2,\,\bar{Q}^1_1,\,\bar{Q}^1_2,\,\bar{Q}^2_1,\, 
\bar{Q}^2_2\}\leftrightarrow\{b_1^\dagger,\,b_2^\dagger,\,b_3^\dagger,\, 
b_4^\dagger,\,b_5^\dagger,\,b_6^\dagger,\,b_7^\dagger,\,b_8^\dagger\} 
\eea

The same oscillator superalgebra can be used to realise the gl$(8|8)$ Lie superalgebra\footnote{For a related discussion, in the setting of nuclear physics, see \cite{BahaBalantekin:1981kt}.}. After imposing a hermiticity condition, using the standard super-oscillator inner product, we obtain u$(8|8)$. The oscillator commutation relations are
\bea
[a_I,a_J^\dagger]\,\,=\,\,\delta_{IJ} \qquad \{b_I,b_J^\dagger\}\,\,=\,\,\delta_{IJ} 
\eea
where $I,J\in\{1,2,\cdots,8\}$. Define gl$(8|8)$ generators as follows
\bea 
&& E_{IJ}\,\,=\,\,a_I^\dagger a_J \qquad \bar E_{IJ}\,\,=\,\,b_I^\dagger b_J\cr\cr
&& F_{IJ}\,\,=\,\,a_I^\dagger b_J \qquad \bar F_{IJ}\,\,=\,\,b_I^\dagger a_J 
\label{homogeneousgenerators}
\eea
Any state in the Fock space can be written as a polynomial in the creation operators acting on the Fock vacuum. We can construct a graded basis whose states are obtained by acting with polynomials of a definite degree. Since the generators given in (\ref{homogeneousgenerators}) are the product of a creation operator with an annihilation operator, they do not change the degree of states in the graded basis. In this sense u$(8|8)$ is a homogeneous symmetry. In contrast, from inspection of  (\ref{psu224generators}) it is clear that the psu$(2,2|4)$ is inhomogeneous. The generators close the algebra
\bea 
&& [E_{IJ},E_{KL}]\,\,=\,\,\delta_{JK}E_{IL}-\delta_{LI}E_{KJ}\cr\cr
&& [E_{IJ},F_{KL}]\,\,=\,\,\delta_{JK}F_{IL}\cr\cr
&& [\bar E_{IJ},\bar{E}_{KL}]\,\,=\,\,\delta_{JK}\bar{E}_{IL}-\delta_{LI}\bar{E}_{KJ}\cr\cr
&& \{\bar{F}_{KL},F_{IJ}\}\,\,=\,\,\delta_{JK}E_{IL}+\delta_{IL}\bar{E}_{KJ}
\eea
which is easily demonstrated by making use of the identities
\bea\label{usefulid}  
&& [ AB , CD ] = A[B,C]D + AC [B,D] + [A,C] DB + C [A,D]B \cr\cr
&&  \{AB,CD\} = A [ B , C ] D + AC [ B , D ] + \{ A , C \} DB + C [ D , A ]B
\eea

Given polynomials $P,Q$ in the creation oscillators, we define an inner product as
\bea 
(P,Q)\,\,=\,\,\langle 0 | P^{ \dagger} Q |0 \rangle \label{InnerProduct}
\eea
where $I,J\in\{1,2,\cdots,8\}$. As usual, the annihilation operators annihilate the Fock vacuum. The dagger of a product of oscillators is the product written in reverse order, with the dagger of each individual oscillator. In particular, there are no minus signs introduced when we take the dagger of a product which includes fermionic oscillators. With respect to this inner product, the hermitian generators of u$(8|8)$ are
\bea 
\begin{array}{ccc}
E_{II} & &\\
E_{IJ} + E_{JI}  & \hbox{for} & I< J\\
i (E_{IJ} - E_{JI}) & \hbox{for}  & I<J  \\
\bar E_{II} & & \\
\bar{E}_{IJ}+\bar{E}_{JI}  & \hbox{for} & I < J \\
i (E_{IJ} - E_{JI})  & \hbox {for} & I<J \\
F_{IJ} + \bar{F}_{JI} & &\\
i F_{IJ} - i\bar{F}_{JI}  &  &
\end{array}  
\eea
where $I,J\in\{1,2,\cdots,8\}$.

\subsection{Oscillator inner product and syzygies}\label{syzygyconstruct} 

This subsection gives a physical algorithm for constructing the syzygies using the oscillator Fock spaces, with the natural inner product (\ref{InnerProduct}). As explained in Section \ref{RingBackground}, the image in ${\cal V}^{(d-1)}_\cR$ of certain linear combinations of generators of a module ${\cal V}^{(d)}_\cR$, vanish. We will denote this linear combination by $g^{(i),[d]}_a$ and talk about a relation obeyed by the image of the generators. Each of these linear combinations defines a syzygy and each syzygy generates a module. In our oscillator Fock space language it is natural to associate a ground state $|0;g^{(i),[d]}_a\rangle$ to each linear combination $g^{(i),[d]}_a$. The direct sum of the Fock spaces built on each of these states is the syzygy module ${\cal V}^{(d+1)}_\cR$. The image of this syzygy module give the kernel of the map from ${\cal V}^{(d)}_\cR$ to ${\cal V}^{(d-1)}_\cR$. The key idea of our algorithm is the fact that this kernel can be determined as the null space of the image of a basis of ${\cal V}^{(d)}_\cR$. The analysis is performed at a given degree, so that all null spaces are finite dimensional. The quadratic relations that we impose on the image of the ring ${\cal V}_{\cR}^{(0)}$, to recover the psu$(2,2|4)$ irreducible representation $V^{\rm phys}$ correspond to the first syzygies and the module ${\cal V}_{\cR}^{(1)}$ is a direct sum of the Fock spaces constructed on the ground states $|0;g^{(1),[d]}_a\rangle$. The relations between the images of the first syzygies define the second syzygies, and the relations between the image of the second syzygies define the third syzygies and so on. The syzygy construction algorithm obtains the syzygies at depth $i+1$ from the syzygies depth $i$. Recall that the depth $i+1$ syzygies are written as linear combinations of the depth $i$ syzygies as follows
\bea
g^{(i+1),[d]}_a=\sum_{j,d'} c^{[d-d']}_{j,a} g^{(i),[d']}_j
\eea
where the coefficients $c^{[d-d']}_j\in {\cal V}_{\cR}^{(0)}$ are of degree $d-d'$. In the oscillator language the coefficients $c^{[d-d']}_j$ are symmetric monomials of degree $d-d'$ in the oscillators and $j$ runs over a basis of (graded) symmetric tensors. The $g^{(i),[d']}_j$ translate into distinct ground states $|0;g^{(i),[d']}_j\rangle$ in a Fock space and the $g^{(i+1),[d]}_a$ become states $|g^{(i+1),[d]}_a\rangle$ in this Fock space
\bea 
|g^{(i+1),[d]}_a\rangle =\sum_{j,[d']} c^{[d-d']}_j |0;g^{(i),[d']}_j\rangle
\eea
We can consider excitations of the states $|g^{(i+1),[d]}_j\rangle$, obtained by acting on the state with creation operators. Introduce the notation
\bea
|{\cal E},[\tilde{d}];\alpha\rangle =T_\alpha^{i_1 i_2\cdots i_{\tilde{d}-d}}a_{i_1}^\dagger\cdots b_{i_{\tilde{d}-d}}^\dagger
|g^{(i+1),[d]}_b\rangle
\eea
for the collection of all excitations of a given degree $\tilde{d}-d$. Here $\alpha$, which is a state label, runs over a basis for graded symmetric tensors. A basis for these excitations is constructed by acting on the ground states $|0;g^{(i),[d']}_b\rangle$. We can again grade the basis by degree. Introduce the notation
\bea
|{\cal B},[\tilde{d}];\beta\rangle = T_\beta^{i_1 i_2\cdots i_{\tilde{d}-d}}a_{i_1}^\dagger\cdots b_{i_{\tilde{d}-d'}}^\dagger
|0;g^{(i),[d']}_b\rangle
\eea
where $\beta$ is a state label. The syzygy construction algorithm considers the matrix of inner products
\bea
 M^{[\tilde{d}]}_{\alpha\beta}
=\langle{\cal E},[\tilde{d}];\alpha|{\cal B},[\tilde{d}];\beta\rangle
\eea
Since the syzygies at depth $i+1$ correspond to relations between the syzygies at depth $i$, the null vectors of $M^{[\tilde{d}]}_{\alpha\beta}$ give the syzygies at depth $i+1$ and their excitations. More precisely, the syzygies are given by the left null space of $M^{[\tilde{d}]}_{\alpha\beta}$, i.e. the set of vectors $v^{(a)}_\alpha$ such that
\bea
\sum_{\alpha}v^{(a)}_\alpha M^{[\tilde{d}]}_{\alpha\beta}=0
\eea

The resolution of the fundamental representation of su(3) provides a nice illustration of the algorithm. The relevant ring and resolution was described in detail in Section \ref{RingsfromLieAlgebras}. Recall that the relevant ring has two generators $x_{21},x_{31}$ and the constraints that must be imposed (see equation (\ref{deff1f2f3})) on the ring to recover su(3) are 
\bea 
g^{(1),[2]}_1 = x_{21}^2 \qquad g^{(1),[2]}_2 = x_{21} x_{31} \qquad 
g^{(1),[2]}_3 = x_{31}^2 
\eea
These are the first syzygies.  In the oscillator language, we have two bosonic oscillators $a_1^\dagger,a_2^\dagger$ which are related to the polynomials as $x_{21}\to a_1^\dagger$, $x_{31}\to a_2^\dagger$. There is a map $\varphi_{10}:{\cal V}^{(1)}_\cR\to{\cal V}_\cR^{(0)}$ and the image of the first syzygies under this map is as follows
\bea
\varphi_{10}(|g^{(1),[2]}_1\rangle) =a_1^\dagger a_1^\dagger |0\rangle\qquad\quad
\varphi_{10}(|g^{(1),[2]}_2\rangle) =a_1^\dagger a_2^\dagger |0\rangle\qquad\quad
\varphi_{10}(|g^{(1),[2]}_3\rangle) =a_2^\dagger a_2^\dagger |0\rangle
\eea
and the excited states of degree 3 are given by
\bea
|{\cal E},[3];1\rangle=a_1^\dagger|g^{(1),[2]}_1\rangle\qquad
|{\cal E},[3];2\rangle=a_2^\dagger|g^{(1),[2]}_1\rangle\qquad
|{\cal E},[3];3\rangle=a_1^\dagger|g^{(1),[2]}_2\rangle\cr\cr
|{\cal E},[3];4\rangle=a_2^\dagger|g^{(1),[2]}_2\rangle\qquad
|{\cal E},[3];5\rangle=a_1^\dagger|g^{(1),[2]}_3\rangle\qquad
|{\cal E},[3];6\rangle=a_2^\dagger|g^{(1),[2]}_3\rangle
\eea
Each of the $|{\cal E},[3];i\rangle$ is given by a degree 3 polynomial in the creation operators, acting on the Fock vacuum. Consequently a basis $|{\cal B},\beta\rangle$ is provided by
\bea
&&|{\cal B},[3];1\rangle=(a_1^\dagger)^3|0\rangle\qquad
|{\cal B},[3];2\rangle=(a_1^\dagger)^2a_2^\dagger|0\rangle\cr\cr
&&|{\cal B},[3];3\rangle=a_1^\dagger(a_2^\dagger)^2|0\rangle\qquad
|{\cal B},[3];4\rangle=(a_2^\dagger)^3|0\rangle\nonumber
\eea
Next, we construct the matrix of oscillator inner products 
\bea 
\varphi_{10}(|{\cal E},[3];\alpha\rangle)^\dagger |{\cal B},[3];\beta\rangle=
\begin{pmatrix} 6 & 0 & 0 & 0 \\ 0 & 2 &  0 & 0 \\ 0 & 0 & 2 & 0 \\ 0 & 2 & 0 & 0 \\ 0 & 0 & 2 & 0 \\ 0 & 0 & 0 & 6 \end{pmatrix}_{\alpha\beta}  
\eea
The null space of this matrix can be calculated in Sage (using left\_kernel)  or in mathematica (using the NullSpace command). There are two null linear combinations, given as 
\bea 
|g^{(2),[3]}_1\rangle=a_1^\dagger |g^{(1),[2]}_2\rangle 
- a_2^\dagger |g^{(1),[2]}_1\rangle\qquad\qquad
|g^{(2),[3]}_2\rangle=a_1^\dagger |g^{(1),[2]}_3\rangle 
- a_2^\dagger |g^{(1),[2]}_2\rangle 
\eea
Translating this back into the polynomial language we have
\bea
g^{(2),[3]}_1=-x_{21}g^{(1),[2]}_2+x_{31}g^{(1),[2]}_3\qquad 
g^{(2),[3]}_2=-x_{21}g^{(1),[2]}_3+x_{31}g^{(1),[2]}_1
\eea
which reproduces the second syzygies given in (\ref{2ndfundsyz}). This accounts for all of the syzygies in the resolution. One can look for second syzygies of degree 4. The excited states for this computation are of degree 4 and are obtained by acting with two creation operators on the first syzygy state. An example of an excited state is
\bea
|{\cal E},[4];1\rangle=a_1^\dagger a_1^\dagger|g^{(1),[2]}_1\rangle
\eea
There are a total of 9 excited states obtained in this way. The elements of the basis are given by acting with four creation operators on the Fock vacuum. An example of a basis state is
\bea
|{\cal B},[4];1\rangle=(a_1^\dagger)^4|0\rangle\qquad
\eea
There are a total 5 states in this basis. Consequently $M^{[4]}_{\alpha\beta}$ has 9 rows and 5 columns. The null space of $M^{[4]}_{\alpha\beta}$ is 4 dimensional. This null space will contain excitations of the two states $|g^{(1),[2]}_1\rangle$, $|g^{(1),[2]}_2\rangle$ discovered at degree 3, as well potentially new syzygies. Applying the operator
\bea
P&=&\mathbb{I}
-\sum_{a,b=1}^2 a_b^\dagger |g^{(2),[3]}_a\rangle\langle g^{(2),[3]}_a|a_b\cr\cr
&=&\sum_{\alpha=1}^5|{\cal B},[4];\alpha\rangle\langle{\cal B},[4];\alpha|
-\sum_{a,b=1}^2 a_b^\dagger |g^{(2),[3]}_a\rangle\langle g^{(2),[3]}_a|a_b
\eea
to the states in the null space of $M^{[4]}_{\alpha\beta}$, we project out the excitations of the known degree three syzygies, leaving only genuinely new syzygies. For the example considered here the projector $P$ annihilates the null space, proving that there are no second syzygies of degree 4. 

Apart from second syzygies of higher degree, one can also search for third syzygies. There are two second syzygies, defining the two ground states on which the second syzygy module is constructed. Each ground state is annihilated by both annihilation operators
\bea
 a_1|0;g^{(1),[2]}_i\rangle = a_2|0;g^{(1),[2]}_i\rangle \qquad i=1,2,3
\eea
Under the map $\varphi_{21}:{\cal V}^{(2)}\to{\cal V}^{(1)}$ the image of the second syzygies are
\bea 
\varphi_{21}(|g^{(2),[3]}_1\rangle)&=&a_1^\dagger |0;g^{(1),[2]}_2\rangle 
- a_2^\dagger |0;g^{(1),[2]}_1\rangle\cr\cr
\varphi_{21}(|g^{(2),[3]}_2\rangle)&=&a_1^\dagger |0;g^{(1),[2]}_3\rangle 
- a_2^\dagger |0;g^{(1),[2]}_2\rangle
\eea
Arguing exactly as we did above, we consider the excited states
\bea
|{\cal E},[4];1\rangle=a_1^\dagger|g^{(2),[3]}_1\rangle\qquad
|{\cal E},[4];2\rangle=a_2^\dagger|g^{(2),[3]}_1\rangle\cr\cr
|{\cal E},[4];3\rangle=a_1^\dagger|g^{(2),[3]}_2\rangle\qquad
|{\cal E},[4];4\rangle=a_2^\dagger|g^{(2),[3]}_2\rangle
\eea
and the basis $|{\cal B},\beta\rangle$ gived by
\bea
|{\cal B},[4];1\rangle=(a_1^\dagger)^2|0;g^{(1),[2]}_1\rangle\qquad
|{\cal B},[4];2\rangle=a_1^\dagger a_2^\dagger|0;g^{(1),[2]}_1\rangle\qquad
|{\cal B},[4];3\rangle=(a_2^\dagger)^2|0;g^{(1),[2]}_1\rangle\cr\cr
|{\cal B},[4];4\rangle=(a_1^\dagger)^2|0;g^{(1),[2]}_2\rangle\qquad
|{\cal B},[4];5\rangle=a_1^\dagger a_2^\dagger|0;g^{(1),[2]}_2\rangle\qquad
|{\cal B},[4];6\rangle=(a_2^\dagger)^2|0;g^{(1),[2]}_2\rangle\cr\cr
|{\cal B},[4];7\rangle=(a_1^\dagger)^2|0;g^{(1),[2]}_3\rangle\qquad
|{\cal B},[4];8\rangle=a_1^\dagger a_2^\dagger|0;g^{(1),[2]}_3\rangle\qquad
|{\cal B},[4];9\rangle=(a_2^\dagger)^2|0;g^{(1),[2]}_3\rangle\cr
\eea
The matrix $M^{[4]}_{\alpha\beta}=\varphi_{21}(|{\cal E},[4];\alpha\rangle)^\dagger|{\cal B},[4];\beta\rangle$ has no null states, indicating that there are no third syzygies of degree 4, which is the correct answer. As a consequence of Hilbert's syzygy theorem we know that there are no higher syzygies, which implies that the corresponding matrices $M^{[d]}_{\alpha\beta}$ constructed using the second (for the $|{\cal E},[d];a\rangle$ vectors) and third syzygies (for the $|{\cal B},[d];b\rangle$ vectors) have no left null vectors.

We have illustrated our algorithm to obtain the syzygies of a resolution using commutative rings. The same algorithm holds for super-rings. An instructive simple example is provided by a super-ring with two commuting coordinates ($x_1$, $x_2$) and two anticommuting coordinates ($q_1$, $q_2$). We impose the following 4 first syzygies
\bea
g^{(1),[2]}_1=x_1^2\qquad g^{(1),[2]}_2=x_1 x_2\qquad g^{(1),[2]}_3=x_2^2
\qquad g^{(1),[2]}_4=q_1q_2
\eea
There are a total of 7 second syzygies, given by
\bea
&&g^{(2),[3]}_1=q_2 g^{(1),[2]}_4\qquad g^{(2),[3]}_2=q_1 g^{(1),[2]}_4\qquad
g^{(2),[3]}_3=x_1 g^{(1),[2]}_2-x_2 g^{(1),[2]}_1\cr\cr
&&g^{(2),[3]}_4=x_1g^{(1),[2]}_3-x_2g^{(1),[2]}_2\qquad\qquad
g^{(2),[4]}_5=x_2^2g^{(1),[2]}_4-q_1q_2 g^{(1),[2]}_3\cr\cr
&&g^{(2),[4]}_6=x_1x_2 g^{(1),[2]}_4-q_1q_2 g^{(1),[2]}_2\qquad
g^{(2),[4]}_7=x_1^2 g^{(1),[2]}_4-q_1q_2 g^{(1),[2]}_1
\eea
Notice that there are second syzygies of both degree 3 and degree 4. There are a total of 11 third syzygies, given by
\bea
   &&g^{(3),[4]}_1=q_ 2 g^{(2),[3]}_1\qquad \qquad
       g^{(3),[4]}_2=q_ 1g^{(2),[3]}_1+q_2g^{(2),[3]}_2\qquad
    \qquad g^{(3),[4]}_3=q_1g^{(2),[3]}_2\cr\cr
   &&g^{(3),[5]}_4=q_1q_2g^{(2),[3]}_3+x_1g^{(2),[4]}_6-x_2g^{(2),[4]}_7\qquad 
       g^{(3),[5]}_5=q_1q_2g^{(2),[3]}_4+x_1g^{(2),[4]}_5-x_2g^{(2),[4]}_6\cr\cr
   &&g^{(3),[5]}_6=x_2^2g^{(2),[3]}-q_2g^{(2),[4]}_5\qquad
       g^{(3),[5]}_7=x_2^2g^{(2),[3]}-q_1g^{(2),[4]}_5\cr\cr
   &&g^{(3),[5]}_8=x_1x_2g^{(2),[3]}_1-q_2g^{(2),[4]}_6\qquad
       g^{(3),[5]}_9=x_1x_2g^{(2),[3]}_2-q_1g^{(2),[4]}_6\cr\cr
   &&g^{(3),[5]}_{10}=x_1^2g^{(2),[3]}_1-q_2g^{(2),[4]}_7\qquad
       g^{(3),[5]}_{11}=x_1^2g^{(2),[3]}_2-q_1g^{(2),[4]}_7
\eea
The third syzygies have degree 4 and degree 5. The resolution is completed with 17 fourth syzygies, given by
\bea
&& g^{(4),[5]}_1=q_2g^{(3),[4]}_1\qquad  
     g^{(4),[5]}_2=q_1g^{(3),[4]}_1+q_2g^{(3),[4]}_2\qquad 
     g^{(4),[5]}_3=q_1g^{(3),[4]}_2+q_2g^{(3),[4]}_3\cr\cr
&& g^{(4),[5]}_4=q_1g^{(3),[4]}_3\qquad
    g^{(4),[6]}_5=x_1g^{(3),[5]}_6-x_2g^{(3),[5]}_8+q_2g^{(3),[5]}_5\cr\cr
&&g^{(4),[6]}_6=x_1g^{(3),[5]}_7-x_2g^{(3),[5]}_9+q_1g^{(3),[5]}_5\qquad
    g^{(4),[6]}_7=x_1g^{(3),[5]}_8-x_2g^{(3),[5]}_{10}+q_2g^{(3),[4]}_4\cr\cr
&&g^{(4),[6]}_8=x_1g^{(3),[5]}_9-x_2g^{(3),[5]}_{11}+q_1g^{(3),[4]}_4\qquad
    g^{(4),[6]}_9=x_2^2g^{(3),[4]}_1-q_2g^{(3),[5]}_6\cr\cr
&&g^{(4),[6]}_{10}=x_2^2g^{(3),[4]}_2-q_1g^{(3),[5]}_6-q_2g^{(3),[5]}_7\qquad
    g^{(4),[6]}_{11}=x_2^2g^{(3),[4]}_3-q_1g^{(3),[5]}_7\cr\cr
&&g^{(4),[6]}_{12}=x_1x_2g^{(3),[4]}_1-q_2g^{(3),[5]}_8\qquad
    g^{(4),[6]}_{13}=x_1x_2g^{(3),[4]}_2-q_1g^{(3),[5]}_8-q_2g^{(3),[5]}_9\cr\cr
&&g^{(4),[6]}_{14}=x_1x_2g^{(3),[4]}_3-q_1g^{(3),[5]}_9\qquad
    g^{(4),[6]}_{15}=x_1^2g^{(3),[4]}_1-q_2g^{(3),[5]}_{10}\cr\cr
&&g^{(4),[6]}_{16}=x_1^2g^{(3),[4]}_2-q_1g^{(3),[5]}_{10}-q_2g^{(3),[5]}_{11}
\qquad
    g^{(4),[6]}_{17}=x_1^2g^{(3),[4]}_3-q_1g^{(3),[5]}_{11}
\eea
These are of degree 5 and degree 6. The Fock space description uses four oscillators, two bosonic oscillators $a_1^\dagger$, $a_2^\dagger$ and two fermionic oscillators $b_1^\dagger$, $b_2^\dagger$. We have checked that the syzygy construction algorithm reproduces this resolution in complete detail.

\subsection{ Orthogonal decomposition of (super-)ring from the resolution of Lie algebra representation  }

The extra structure provided by the u$(8|8)$ inner product of the Fock space allows us to
explain how the resolution of an irrep $V$ specifies an orthogonal decomposition for ${\cal V}^{(0)}$ into a direct sum of $V$ and its orthogonal complement. To make the discussion concrete, consider the resolution of an irrep $V$ given by the following exact sequence
\bea
\begin{tikzcd}
  0 \arrow[r," "] & {\cal V}^{(3)}  \arrow[r, "\varphi_{32}"] & {\cal V}^{(2)}\arrow[r, "\varphi_{21}"] & {\cal V}^{(1)}\arrow[r,"\varphi_{10}"] & {\cal V}^{(0)}\arrow[r,"\varphi_{0*}"]& V \arrow[r," "] & 0
\end{tikzcd}
\nonumber
\eea
Making use of the discussion given in Section \ref{RingsfromLieAlgebras} above, we know that the resolution implies the following short exact sequences
\bea
&&
\begin{tikzcd}
  0 \arrow[r," "] & {\rm Im}(\varphi_{10})  \arrow[r, " "] & {\cal V}^{(0)} \arrow[r, " "] &V \qquad\,\,\,\arrow[r," "] & 0
\end{tikzcd}\cr
&&
\begin{tikzcd}
  0 \arrow[r," "] & {\rm Im}(\varphi_{21})  \arrow[r, " "] & {\cal V}^{(1)} \arrow[r, " "] &{\rm Im}(\varphi_{10}) \arrow[r," "] & 0
\end{tikzcd}\cr
&&
\begin{tikzcd}
  0 \arrow[r," "] & {\rm Im}(\varphi_{32})  \arrow[r, " "] & {\cal V}^{(2)} \arrow[r, " "] &{\rm Im}(\varphi_{21}) \arrow[r," "] & 0
\end{tikzcd}\cr
&&
\begin{tikzcd}
 &\qquad\qquad 0 \,\, \arrow[r, " "] & {\cal V}^{(3)} \arrow[r, " "] &{\rm Im}(\varphi_{32}) \arrow[r," "] & 0
\end{tikzcd}
\eea
Using the oscillator inner product, these short exact sequences imply the following orthogonal decompositions
\bea
{\cal V}^{(0)}&=&V\oplus {\rm Im}(\varphi_{10})\label{1a}
\eea
\bea
{\cal V}^{(1)}&=&{\rm Im}(\varphi_{10})\oplus {\rm Im}(\varphi_{21})\label{1b}
\eea
\bea
{\cal V}^{(2)}&=&{\rm Im}(\varphi_{21})\oplus {\rm Im}(\varphi_{32})\label{1c}
\eea
In what follows we will denote the subspace of ${\cal V}$ that is orthogonal to $A$ by $(A)_{\perp ; {\cal V}}$. From (\ref{1c}) we have
\bea
{\rm Im}(\varphi_{21})=({\rm Im}(\varphi_{32}))_{\perp ; {\cal V}^{(2)}}
\eea 
while (\ref{1b}) implies that
\bea
{\rm Im}(\varphi_{10})=({\rm Im}(\varphi_{21}))_{\perp ; {\cal V}^{(1)}}
\eea 
Taken together we have
\bea
{\rm Im}(\varphi_{10})=((({\rm Im}(\varphi_{32}))_{\perp ; {\cal V}^{(2)}})_{\perp ; {\cal V}^{(1)}}
\eea 
Finally, using (\ref{1a}) we find
\bea
V=({\rm Im}(\varphi_{10}))_{\perp ; {\cal V}^{(0)}}
=(((({\rm Im}(\varphi_{32}))_{\perp ; {\cal V}^{(2)}})_{\perp ; {\cal V}^{(1)}})_{\perp ; {\cal V}^{(0)}}
\eea
If we recall that ${\cal V}^{(3)}={\rm Im}(\varphi_{32})$, this can also be written as
\bea
V=((({\cal V}^{(3)})_{\perp ; {\cal V}^{(2)}})_{\perp ; {\cal V}^{(1)}})_{\perp ; {\cal V}^{(0)}}
\eea
or, equivalently
\bea
{\cal V}^{(0)}=V\oplus
(({\cal V}^{(3)})_{\perp ; {\cal V}^{(2)}})_{\perp ; {\cal V}^{(1)}}
\eea
Thus, the exact sequence together with the oscillator inner product allows us to give a description of ${\cal V}^{(0)}$ as a direct sum of $V$ and its orthogonal complement. Using only (\ref{1a}) together with the oscillator inner product, we obtain
\bea
V=({\rm Im}(\varphi_{10}))_{\perp ; {\cal V}^{(0)}}
\eea
which gives a description of $V$ as a subspace of ${\cal V}^{(0)}$. This gives a basis for $V$, but ${\rm Im}(\varphi_{10})$ is described using the explicit basis for ${\cal V}^{(1)}$ together with the map $\varphi_{10}$. This map has a non-trivial kernel, so that we have not obtained a basis for ${\rm Im}(\varphi_{10})$. Making use of the complete resolution and the oscillator inner product as above, gives a $\mathbb{C}$-basis for $V$ and its orthogonal complement.

\subsection{ Number operator Hamiltonian as anti-commutator of supercharges } 

The syzygy construction algorithm outlined in Section \ref{syzygyconstruct} employs a Fock space and represents syzygies as states in this Fock space. This allows us to translate results from the language of rings and modules into the language of quantum mechanics. In this section we develop this point of view further. To start, define the operators
\bea 
Q\,\, =\,\,\sum_{I,J=1}^8 M^{IJ} F_{IJ} \qquad \bar{Q}\,\, =\,\,\sum_{K,L=1}^8 M^{KL} \bar{F}_{KL} 
\eea
which obey
\bea 
\{ \bar Q  , Q \} \,\,=\,\, \sum_{K,L,J=1}^8 M^{KL}M^{LJ}\bar{E}_{KJ} 
+\sum_{I,K,L=1}^8 M^{KL} M^{IK} E_{IL}  
\eea
If we now choose $M M^T = 1$, then
\bea 
\{\bar{Q},Q\}=\sum_{J=1}^8\bar{E}_{JJ}+\sum_{I=1}^8 E_{II} 
\eea
The right hand side is the number operator. Thus, we have a super-quantum mechanics with a Hamiltonian equal to the number operator. This Hamiltonian is the anti-commutator of two super-charges. 

The state space of this quantum mechanical system admits an action of psu$(2,2|4)$, realized as the differential operators given in (\ref{psu224generators}). While the u$(8|8)$ symmetry takes states of a fixed degree to other states of the same degree, the action of the psu$(2,2|4)$ symmetry does not preserve the degree: it causes a weak mixing of degrees. 

The ideal $I$ - given by the constraints - is a representation of psu$(2,2|4)$ but not of u$(8|8)$. The quotient, which is the half BPS irrep, is also a representation of psu$(2,2|4)$ but not of u$(8|8)$. So this is an interesting example of an ideal constructed from oscillators leading to a new symmetry on the quotient. 

\section{ Summary and future directions } \label{summaryanddirections}

We give  a brief summary of our main results. Building on the well-known BGG resolution \cite{BGG} of lowest weight irreps $V_{LW}$ of Lie algebras $ \mg$,  and motivated by the structure of half-BPS representations of the psu$(2,2|4)$ symmetry of $ \cN =4 $ super Yang Mills theory, we have explained a variation of the BGG resolution. Lowest weight representations are defined using a decomposition of the Lie algebra $\mg$ associated with positive and negative  roots and the Cartan sub-algebra : $ \mg = \mg^{+} \oplus \mg^{0} \oplus \mg^{-} $.  BGG resolutions give a description of a $V_{LW}$ as the last term in an exact sequence of $U(\mg)$  modules. The terms to the left of $V_{LW}$ are Verma modules obtained by acting with $U (\mg^+) $ on a lowest weight state.  These resolutions are known to  be $U(\mg)$ equivariant \cite{BGG}. In the construction we presented, $V_{LW} $ arises as the right-most term of an exact sequence of modules of $ U (\mg^+_c) $, where $ \mg_c^+$ is a commutative Lie sub-algebra of $\mg^{+}$. The enveloping algebra $U ( \mg_c^+)= \cR $ is a (super-)polynomial ring in the case of Lie (super-)algebras.  In a number of examples, including the psu$(2,2|4)$ half-BPS representation, we showed that $\cR = U ( \mg_c^+) $ is also a representation of $U ( \mg )$. The whole sequence is generically not $U(\mg)$ equivariant, but enjoys the property that it is amenable to standard techniques for computations with polynomial (super-)rings, as available in SageMath for example \cite{Cox}.  For the case of  half-BPS  psu$(2,2|4)$ representations where the lowest weight state corresponds to $\tr Z^n $, $\mg_c^+$ is a commutative Lie superalgebra with eight bosonic and eight fermionic generators and we denote $U ( \mg_c^+ ) = \cR ( 8 |8  )$. Associated with $\mg_c^+$  there is a super-oscillator algebra, with one creation operator for every generator of $\mg_c^+$. The ring of (super-)polynomials in the creation operators is isomorphic to $U ( \mg_c^+ ) = \cR ( 8 |8  )$. 
There is an algebra  u$(8|8)$ Lie superalgebra which  can be constructed from the creation and annihilation operators of the oscillator algebra. This algebra acts homogeneously on the Fock space states, with basis given by the action of $\cR( 8 |8)$ on a ground state, i.e. when acting on states with a fixed number of creation operators, they preserve this number. The oscillators can also be used to construct the generators of  $\psu(2,2|4)$ which act on $U(\mg_c^+)=\cR(8|8)$ or equivalently on the Fock space states. The action of $\psu(2,2|4)$ is not homogeneous, i.e. does not preserve the number of creation operators. 
There is a  u$(8|8)$ compatible inner product on $\cR ( 8 |8 )$ which  is useful in giving a physical understanding (see Section \ref{syzygyconstruct}) of the algorithms for constructing the exact sequence of $ \cR ( 8|8 ) $ modules which resolves the half-BPS irrep $V_{\rm phys}^{\rm BPS}$. 

A number of interesting future research directions follow from the above results. The $U(1)$ $\cN=4$ super-Yang Mills theory is the theory on a single D3-brane in type IIB string theory. The three- and five-dimensional theories on a single M2-brane and a single M5-brane have the same amount of supersymmetry as the D3-brane theory \cite{Minwalla:1998rp}. It will be interesting to develop an understanding of resolutions of the fundamental field multiplet of the single M2- and M5- theories in terms of super-polynomial rings, analogous to the discussion here. 

A fascinating direction is to better understand the physics of the u$(8|8)$ symmetry that has appeared in the description of a resolution of the half-BPS representation of four-dimensional  $U(1)$ $\cN=4$ super-Yang Mills. This symmetry, hidden from the point of view of $\cN=4$ SYM theory, may be  related to the fact that the four-dimensional theory arises as the dimensional reduction of ten dimensional super-Yang Mills \cite{HS,Movshev:2003ib}. It may also provide hints towards a constructive avenue, e.g along the lines of collective field theory \cite{Jevicki:1979mb,Jevicki:1980zg},  for the understanding of the emergence of the higher dimensional AdS/CFT dual of four-dimensional $\cN=4$ SYM theory. Unravelling the implications of the u$(8|8)$ for the physical observables in AdS/CFT would be a good step in this direction. With the polynomial super-ring description for the fundamental field representation of $U(1)$ SYM and for the $\tr Z$ multiplet of $U(N)$ SYM in hand, the next step is to find super-ring descriptions for generic multiplets in $U(N)$ $\cN=4$ SYM theory. 

From the oscillator description we discussed in Section \ref{emergentsymm}, u$(8|8)$ is the more manifest symmetry
of the $\cR(8|8)$ ring generated by  eight bosonic and eight fermionic creation operators. From this starting point the existence of the exact sequence  (\ref{resHalfBPS}) of $\cR ( 8 |8)$ modules terminating in $ V^{\rm{BPS}}_{\rm{phys}} $ is the origin of the emergence of psu$(2,2|4)$ symmetry. Algebraic mechanisms for the emergence of symmetries, notably Schur-Weyl duality and the double-commutant theorem, continue to  play a significant role in holography, many-body quantum physics and quantum information \cite{SWinst,BM1702,DHKZ2005,BPRQM,MoMo2209,horodeki}. It would be fascinating to uncover the wider applicability, in holography and many-body quantum physics,  of the mechanism of emergence based on oscillator systems and associated resolutions of ring modules described in Section \ref{emergentsymm}. 

\begin{center} 
{\bf Acknowledgements}
\end{center} 
SR is supported by the STFC consolidated grant ST/P000754/1 `` String Theory, Gauge Theory and  Duality”.  RdMK is supported by the South African Research Chairs initiative  of the Department of Science and Technology and the National Research Foundation. Some of this work was completed while RdMK was a participant of the KITP program ``Integrability in String, Field, and Condensed Matter Theory.'' This research was also supported in part by the National Science Foundation under Grant No. NSF PHY-1748958. We thank Yang Hui He, Antal Jevicki, Adrian Padellaro and Gabrielle Travaglini for useful discussions  on the subject of this paper.  

\appendix

\section{Conventions}\label{conventions}

For the bispinor notation we frequently use
\begin{eqnarray}
(\sigma^\mu)_{\alpha\dot{\alpha}}=(1, \, \vec{\sigma})\qquad
(\bar{\sigma}^\mu)^{\dot{\alpha}\alpha}=(1, \, -\vec{\sigma})
\end{eqnarray}
where $\vec{\sigma}$ are the Pauli matrices. su$(2)$ indices can be lowered and raised
\begin{eqnarray} 
X^\alpha=\epsilon^{\alpha\beta}X_\beta\qquad X_\alpha=\epsilon_{\alpha\beta}X^\beta
\end{eqnarray}
where $\epsilon_{12}=-1$, $\epsilon^{12}=1$.
The Lorentz generators $M_{\mu\nu}$ split into two adjoint representations, $M_\alpha{}^\beta$ and 
$M^{\dot{\alpha}}{}_{\dot{\beta}}$. 
The momentum and special conformal transformations are bispinors. 
\begin{eqnarray}
P_{\alpha\dot{\alpha}}&=&{(\sigma^\mu)}{_{\alpha\dot{\alpha}}} P_\mu\qquad
K^{\dot{\alpha}\alpha}={(\bar{\sigma}^\mu)}{^{\dot{\alpha}\alpha}} K_\mu\cr
{M}_\alpha{}^\beta&=&-\frac{i}{4} (\sigma^\mu)_{\alpha\dot{\alpha}} (\bar{\sigma}^\nu)^{\dot{\alpha}\beta} 
M_{\mu\nu}\cr
M^{\dot{\alpha}}{}_{\dot{\beta}}&=&-\frac{i}{4} (\bar{\sigma}^\mu)^{\dot{\alpha}\alpha} (\sigma^\nu)_{\alpha\dot{\beta}} M_{\mu\nu}
\end{eqnarray}

\section{The psu$(2,2|4)$ algebra}\label{psu224}

In this Appendix, all Greek indices run over 1,2 all dotted Greek indices run over 1,2 and Latin indices over 1,2,3,4. The conformal algebra is 
\begin{eqnarray}
[M_\alpha{}^\beta,M_\gamma{}^\delta]&=&
\delta_\gamma^\beta M_\alpha{}^\delta-\delta_\alpha^\delta M_\gamma{}^\beta\qquad
[M^{\dot{\alpha}}{}_{\dot{\beta}},M^{\dot{\gamma}}{}_{\dot{\delta}}]\,\,\,=\,\,\,
\delta^{\dot{\gamma}}_{\dot{\beta}} M^{\dot{\alpha}}{}_{\dot{\delta}}
-\delta^{\dot{\alpha}}_{\dot{\delta}} M^{\dot{\gamma}}{}_{\dot{\beta}}\cr\cr
[M_\alpha{}^\beta,P_{\gamma\dot{\gamma}}]
&=&\delta_\gamma^\beta P_{\alpha\dot{\gamma}}
-\frac{1}{2}\delta_\alpha^\beta P_{\gamma\dot{\gamma}}\qquad
[M^{\dot{\alpha}}{}_{\dot{\beta}},P_{\gamma\dot{\gamma}}]\,\,\,=\,\,\,
-\delta^{\dot{\alpha}}_{\dot{\gamma}} P_{\gamma\dot{\beta}}
+\frac{1}{2}\delta^{\dot{\alpha}}_{\dot{\beta}}P_{\gamma\dot{\gamma}}\cr\cr
[M_\alpha{}^\beta,K^{\dot{\gamma}\gamma}]&=&
-\delta_\alpha^\gamma K^{\dot{\gamma}\beta}
+\frac{1}{2}\delta_\alpha^\beta K^{\dot{\gamma}\gamma}\qquad
[M^{\dot\alpha}{}_{\dot\beta},K^{\dot{\gamma}\gamma}]\,\,\,=\,\,\,
\delta^{\dot{\gamma}}_{\dot{\beta}} K^{\dot{\alpha}\gamma}
-\frac{1}{2}\delta^{\dot{\alpha}}_{\dot{\beta}} K^{\dot{\gamma}\gamma}\cr\cr
[K^{\dot{\alpha}\alpha},P_{\beta\dot{\beta}}]&=&
-4\delta_\beta^\alpha \delta^{\dot{\alpha}}_{\dot{\beta}} D
+4\delta_\beta^\alpha M^{\dot{\alpha}}{}_{\dot{\beta}}
-4\delta^{\dot{\alpha}}_{\dot{\beta}} M_\beta{}^\alpha\cr\cr
[D,P_{\alpha\dot{\alpha}}]&=&P_{\alpha\dot{\alpha}}
\qquad
[D,K^{\dot{\alpha}\alpha}]\,\,\,=\,\,\,-K^{\dot{\alpha}\alpha}
\end{eqnarray}
The R-symmetry generators satisfy the algebra
\begin{equation}\label{su4alg}  
[R^i{}_j,R^k{}_l]=-\delta^k_j R^i{}_l+\delta^i_l R^k{}_j
\end{equation}
The commutators involving the supercharges and the conformal supercharges are
\begin{eqnarray}
[M_\alpha{}^\beta,Q^i{}_\gamma]
&=&\delta_\gamma^\beta Q^i{}_\alpha-\frac{1}{2}\delta_\alpha^\beta Q^i{}_\gamma\qquad
[M^{\dot{\alpha}}{}_{\dot{\beta}},\bar{Q}_{i\dot{\gamma}}]\,\,\,=\,\,\,
-\delta^{\dot{\alpha}}_{\dot{\gamma}}\bar{Q}_{i\dot{\beta}}
+\frac{1}{2} \delta^{\dot{\alpha}}_{\dot{\beta}} \bar{Q}_{i\dot{\gamma}}\cr\cr
[M_\alpha{}^\beta,S_i{}^\gamma]&=&-\delta_\alpha^\gamma S_i{}^\beta
+\frac{1}{2} \delta_\alpha^\beta S_i{}^\gamma\qquad
[M^{\dot{\alpha}}{}_{\dot{\beta}},\bar{S}^{i\dot{\gamma}}]\,\,\,=\,\,\,
\delta^{\dot{\gamma}}_{\dot{\beta}} \bar{S}^{i\dot{\alpha}}
-\frac{1}{2}\delta^{\dot{\alpha}}_{\dot{\beta}} \bar{S}^{i\dot{\gamma}}\cr\cr
[R^i{}_j,Q^k{}_\alpha]&=&-\delta_j^k Q^i{}_\alpha+\frac{1}{4}\delta^i_j Q^k{}_\alpha\qquad
[R^i{}_j,\bar{Q}_{k\dot{\alpha}}]\,\,\,=\,\,\,\delta_k^i \bar{Q}_{j\dot{\alpha}}
-\frac{1}{4}\delta^i_j \bar{Q}_{k\dot{\alpha}}\cr\cr
[R^i{}_j, S_k{}^\alpha]&=&\delta^i_k S_j{}^\alpha-\frac{1}{4}\delta^i_j S_k{}^\alpha\qquad
[R^i{}_j, \bar{S}^{k\dot{\alpha}}]\,\,\,=\,\,\,-\delta^k_j \bar{S}^{i\dot{\alpha}}
+\frac{1}{4}\delta^i_j \bar{S}^{k\dot{\alpha}}\cr\cr
[D,Q^i{}_\alpha]&=&\frac{1}{2} Q^i{}_\alpha\qquad
[D,\bar{Q}_{i\dot{\alpha}}]\,\,\,=\,\,\,\frac{1}{2} \bar{Q}_{i{\dot{\alpha}}}\qquad
[D,S_i{}^\alpha]\,\,\,=\,\,\,-\frac{1}{2} S_i{}^\alpha\cr\cr
[D,\bar{S}^i{}_{\dot{\alpha}}]&=&-\frac{1}{2} \bar{S}^i{}_{\dot{\alpha}}\qquad
[P_{\alpha\dot{\alpha}},S_i{}^\beta]\,\,\,=\,\,\,-2\delta_\alpha^\beta \bar{Q}_{i\dot{\alpha}}\qquad
[P_{\alpha\dot{\alpha}},\bar{S}^{i\dot{\beta}}]=2\delta_{\dot{\alpha}}^{\dot{\beta}} Q^i_\alpha\cr\cr
[K^{\dot{\alpha}\alpha},Q^i{}_\beta]&=&2\delta_\beta^\alpha \bar{S}^{i\dot{\alpha}}\qquad
[K^{\dot{\alpha}\alpha},\bar{Q}_{i\dot{\beta}}]\,\,\,=\,\,\,-2\delta_{\dot{\beta}}^{\dot{\alpha}} S_i{}^\alpha
\end{eqnarray}
Finally, the anti-commutators of the supercharges are
\begin{eqnarray}
\{Q^i{}_\alpha,\bar{Q}_{j\dot{\alpha}}\}&=&2\delta^i_j P_{\alpha\dot{\alpha}}
\qquad
\{\bar{S}^{i\dot{\alpha}},S_j{}^\alpha\}\,\,\,=\,\,\,2\delta^i_j K^{\dot{\alpha}\alpha}\cr\cr
\{Q^i{}_\alpha,S_j{}^\beta\}&=&4\delta^i_j M_\alpha{}^\beta+2\delta^i_j \delta_\alpha^\beta D
+4\delta_\alpha^\beta R^i{}_j\cr\cr
\{\bar{S}^{i\dot{\alpha}},\bar{Q}_{j\dot{\beta}}\}&=&4\delta^i_j M^{\dot{\alpha}}{}_{\dot{\beta}}
-2\delta^i_j \delta^{\dot{\alpha}}_{\dot{\beta}} D+4\delta^{\dot{\alpha}}_{\dot{\beta}} R^i{}_j
\end{eqnarray}

\section{Fermionic Fock Spaces}\label{FermionicFock}

Pauli's exclusion principle states that two fermions can not occupy the same state. This can be encoded algebraically, as the statement that fermionic oscillators are Grassmann valued. Consequently, oscillators associated to different species of fermions anticommute and each fermionic oscillator squares to zero. An alternative approach to fermion state spaces is to start with the polynomial ring associated to a collection of bosonic oscillators and to implement Fermi statistics as a set of constraints on this ring. As a result, one obtains a resolution of fermionic Fock spaces in terms of bosonic Fock spaces. In this Appendix we explain the construction.

From experience with free bosonic partition functions we know that the generating function
\bea
\prod_{i=1}^m {1\over 1-x_i}
\eea
counts states in the $m$ boson Fock space, graded on boson species. Each monomial $x_1^{n_1}x_2^{n_2}\cdots x_m^{n_m}$ labelled by $(n_1,n_2,\cdots,n_m)$ which appears once and only once, corresponds to a state
\bea
x_1^{n_1}x_2^{n_2}\cdots x_m^{n_m}\quad\leftrightarrow\quad
(a_1^\dagger)^{n_1}(a_2^\dagger)^{n_2}\cdots (a_m^\dagger)^{n_m}|0\rangle
\eea

Now consider the Fock space associated to $q$-fermions. Our goal is to give a resolution of the fermionic Fock space, with bosonic Fock spaces. The relevant ring is the ring of $q$ bosonic creation operators. The associated Hilbert series is given by the fermionic generating function, which can be written as
\bea
{\rm H}_F=\prod_{i=1}^q (1+y_i)\label{fergenfun}
\eea
We can again put monomials into correspondence with states
\bea
y_1^{n_1}y_2^{n_2}\cdot y_q^{n_q}\quad\leftrightarrow\quad
(b_1^\dagger)^{n_1}(b_2^\dagger)^{n_2}\cdots (b_q^\dagger)^{n_q}|0\rangle
\eea
Expanding ${\rm H}_F$, we find that only occupation numbers $n_i=0,1$ for $i=1,2,\cdots,q$ appear as expected. In addition, setting $y_1=y_2=\cdots=y_q$ we find $H_F=2^q$ which is the correct total number of states in the Fock space.

To construct the resolution, start from the ring of $q$ bosonic oscillators $\mathbb{C}[b_1^\dagger,b_2^\dagger,\cdots,b_q^\dagger]$ and impose the $q$ constraints
\bea
(b_k^\dagger)^2=0\qquad k=1,2,\cdots q
\eea
Using Sage it is straight forward to verify the validity of the construction. For example, if we set $q=3$ and evaluate the code
\begin{verbatim}
names = '(x,y,z)'
Igens = ('x^2','y^2','z^2')
R = singular.ring(0,names,'dp')
I = singular.ideal(Igens)
from sage.rings.polynomial.hilbert import hilbert_poincare_series
hilbert_poincare_series(I)
\end{verbatim}
we obtain the following Hilbert series
\bea
H_F(t)=t^3 + 3t^2 + 3t + 1
\eea
Recall that since each oscillator adds a fermion, the resolution is graded by fermion number. From the above Hilbert series we read off that there is one state of fermion number zero (the Fock space vacuum $|0\rangle$), three states of fermion number one ($b_i^\dagger
|0\rangle$, $i=1,2,3$), three states of fermion number two ($b_i^\dagger b_j^\dagger |0\rangle$, $i<j$ and $i,j=1,2,3$) and one state of fermion number three ($b_1^\dagger b_2^\dagger b_3^\dagger |0\rangle$). The resolution is given by evaluating
\begin{verbatim}
names = '(x,y,z)'
Igens = ('x^2','y^2','z^2')
R = singular.ring(0,names,'dp')
I = singular.ideal(Igens)
singular.res(I,0)
\end{verbatim}
The resolution is
\bea
\begin{tikzcd}
  0 \arrow[r," "] & {\cal V}^{(3)} \arrow[r,"\varphi_{32}"] & {\cal V}^{(2)}  \arrow[r, "\varphi_{21}"] & {\cal V}^{(1)}\arrow[r, "\varphi_{10}"] & {\cal V}^{(0)}\arrow[r,"\varphi_{0*}"] & V^F \arrow[r," "] & 0
\end{tikzcd}
\eea
where ${\cal V}^{(0)}=\mathbb{C}[b_1^\dagger,b_2^\dagger,b_3^\dagger]$, the module ${\cal V}^{(1)}$ is a free module generated by
\bea
f_1=(b_1^\dagger)^2\qquad f_2=(b_2^\dagger)^2\qquad f_3=(b_3^\dagger)^2
\eea
the module ${\cal V}^{(2)}$ is a free module generated by
\bea
g_1=-(b_2^\dagger)^2f_3+(b_3^\dagger)^2f_2\qquad
g_2=-(b_1^\dagger)^2f_3+(b_3^\dagger)^2f_1\qquad
g_3=-(b_1^\dagger)^2f_2+(b_2^\dagger)^2f_1
\eea
and the module ${\cal V}^{(3)}$ is a free module generated by
\bea
h_1=(b_1^\dagger)^2 g_1-(b_2^\dagger)^2 g_2+(b_3^\dagger)^2 g_3
\eea
The module $V^F$ is the Fock space for three species of fermions. There are three oscillators, and there are three syzygy modules, which agrees with Hilbert's Theorem. Using Sage it is simple to verify that for $k$ species of fermions, there are $k$ syzygy modules. 

\section{Sage code for subring resolutions}\label{sagesubring}

\subsection{Antisymmetric of su(4) from $\mathbb{C}[R^3{}_1,R^3{}_2,R^4{}_1,R^4{}_2]$}

In this section we show how to obtain a resolution of the six-dimensional anti-symmetric representation (spanned by $|M_{ij}\rangle$) of the $\mg= su(4)$ Lie algebra (with commutation relations given in \eqref{su4alg}) using the commutative sub-algebra $\mg_c^+$ spanned by $\{ R^3{}_1,R^3{}_2,R^4{}_1,R^4{}_2 \}$. The enveloping algebra $U ( \mg_c^+) $ is the polynomial ring  $\mathbb{C}[R^3{}_1,R^3{}_2,R^4{}_1,R^4{}_2]$. Acting with elements of this ring on the state $|M_{34}\rangle$ produces all the states in the anti-symmetric representation. The sage code
\begin{verbatim}
names = '(r13,r14,r23,r24)'
Igens = ('r13*r13','r13*r23','r23*r23','r14*r14','r14*r24','r24*r24','r13*r14',
'r23*r24','r13*r24+r23*r14')
R = singular.ring(0,names,'dp')
I = singular.ideal(Igens)
from sage.rings.polynomial.hilbert import hilbert_poincare_series
hilbert_poincare_series(I)
\end{verbatim}
produces the Hilbert series
\bea
H(t) = t^2 + 4t + 1
\eea
which is the correct answer, as implied by (\ref{physstates}) which shows that the four states $|M_{13}\rangle$, $|M_{23}\rangle$, $|M_{14}\rangle$ and $|M_{24}\rangle$ are obtained by applying a single su(4) generator to $|M_{34}\rangle$, while the single state $|M_{12}\rangle$ is obtained by applying two generators to $|M_{34}\rangle$. The corresponding resolution is obtained using
\begin{verbatim}
names = '(r13,r14,r23,r24)'
Igens = ('r13*r13','r13*r23','r23*r23','r14*r14','r14*r24','r24*r24','r13*r14',
'r23*r24','r13*r24+r23*r14')
R = singular.ring(0,names,'dp')
I = singular.ideal(Igens)
singular.res(I,0)
\end{verbatim}
This resolution has maximal depth 4 in the notation introduced after (\ref{resHalfBPS}).

\subsection{A boson and a fermion from $\mathbb{C}[R^3{}_1,R^3{}_2,R^4{}_1,R^4{}_2,Q^3,Q^4]$}

In this section we discuss the super subring  of $\cR ( 8|8)$ which leads to a resolution of an irrep given by a boson in the antisymmetric su(4) and a fermion in the fundamental. The six bosonic states of the previous subsection will again contribute $1+4t+t^2$ to the Hilbert series. From the fermion generators we have $|\psi_4\rangle=Q^3|M_{34}\rangle$ and $|\psi_3\rangle=Q^4|M_{34}\rangle$ contributing $2t$ to the Hilbert series, as well as $|\psi_2\rangle=R^4{}_2Q^3|M_{34}\rangle$ and $|\psi_1\rangle=R^4{}_1Q^3|M_{34}\rangle$ contributing $2t^2$. The code
\begin{verbatim}
names = '(r13,r14,r23,r24,q3,q4)'
Igens = ('r13*r13','r13*r23','r23*r23','r14*r14','r14*r24','r24*r24','r13*r14',
             'r23*r24','r13*r24+r23*r14','r13*q3','r23*q3','r14*q4','r24*q4',
             'r14*q3+r13*q4','r24*q3+r23*q4','q3*q4')
R = singular.ring(0,names,'dp')
A = singular.superCommutative(5,6)
A.set_ring()
I = singular.ideal(Igens)
from sage.rings.polynomial.hilbert import hilbert_poincare_series
hilbert_poincare_series(I)
\end{verbatim}
produce the Hilbert series
\bea
H(t) = 3t^2 + 6t + 1
\eea
which is the correct answer. The complete resolution is obtained from
\begin{verbatim}
names = '(r13,r14,r23,r24,q3,q4)'
Igens = ('r13*r13','r13*r23','r23*r23','r14*r14','r14*r24','r24*r24','r13*r14',
             'r23*r24','r13*r24+r23*r14','r13*q3','r23*q3','r14*q4','r24*q4',
             'r14*q3+r13*q4','r24*q3+r23*q4','q3*q4')
R = singular.ring(0,names,'dp')
A = singular.superCommutative(5,6)
A.set_ring()
I = singular.ideal(Igens)
singular.res(I,0)
\end{verbatim}
This resolution has maximal depth 6 in the notation introduced after (\ref{resHalfBPS}).

\end{document}